\documentclass[12pt]{article}
\usepackage{epsfig,axodraw}

\def\vp{\varphi}

\def\a{\alpha}
\def\lan{\langle}
\def\ran{\rangle}

\begin{document}
\title{\bf MHV Vertices and Fermionic Scattering Amplitudes
in Gauge Theory with Quarks and Gluinos}
\author{Jun-Bao Wu \\School of Physics, Peking University \\
Beijing 100871, P. R. China\\ \\
Chuan-Jie Zhu\thanks{Supported in part by fund from the National
Natural Science Foundation of China with grant Number
90103004.} \\
Institute of Theoretical Physics,
Chinese Academy of Sciences\\
P. O. Box 2735,  Beijing 100080, P. R. China}

\maketitle

\begin{abstract}
The Cachazo-Svrcek-Witten approach to perturbative gauge theory is
extended to gauge theories with quarks and gluinos. All googly
amplitudes with quark-antiquark pairs and gluinos are computed and
shown to agree with the previously known results. The computations
of the non-MHV or non-googly amplitudes are also briefly
discussed, in particular the purely fermionic amplitude with 3
quark-antiquark pairs.
\end{abstract}

\section{Introduction}

Recently Witten \cite{Wittenb} found a deep connection between the
perturbative gauge theory and   string theory in twistor space
\cite{Penrosea}. Based on this work, Cachazo, Svrcek and Witten
(CSW for short) reformulated the  perturbative calculation of the
scattering amplitudes in Yang-Mills theory by using the off shell
MHV vertices \cite{Wittena}. The MHV vertices they used are the
usual tree level MHV scattering amplitudes in gauge theory
\cite{Parkea, Giele}, continued off shell in a particular fashion
as given in \cite{Wittena}. (For references on perturbative
calculations, see for example \cite{Parkeb, Berna, Dixon, Bernb,
Others}. The 2 dimensional origin of the MHV amplitudes in gauge
theory was first given in \cite{Nair}.) Some sample calculations
were done in \cite{Wittena}, sometimes with the help of symbolic
manipulation. The correctness of the rules was partially verified
by reproducing the known results for small number of gluons
\cite{Parkeb}.

In two previous works \cite{Zhu,Wu} (for recent works, see
\cite{itp,itpB, Berkovits,BerkovitsMotl,AganagicVafa, Wittenc,
NeitzkeVafa, NekrasovOoguriVafa, GeorgiouKhoze,Gukov, Siegel,
Giombi, Popov,Wittend, Bernc, Kosower, Wittene}), by following the
new approach of \cite{Wittena}, we have computed the generic gluon
googly amplitudes and discussed briefly how the CSW approach can
be extended to theories with fermions (see also
\cite{GeorgiouKhoze}). In this paper  we will present the full set
of the CSW rules for gauge theories with quarks and antiquarks
(fermions in the fundamental representation) and gluinos (in
supersymmetric theory or fermions in the adjoint representation).
Although the fermionic amplitudes at tree level can be   obtained
by supersymmetric Ward identities \cite{Grisarua,Grisarub, Parkeb}
we think it is still worthy to compute these amplitudes directly.
Although the CSW rule can be partially understood from the twistor
string theory \cite{Gukov}, a full understanding of the CSW
approach from the conventional field theory is not reached
\cite{Bernc}.

By using these CSW rules with fermions, we will compute the googly
amplitudes with one and two quark-antiquark pairs. We will also
compute the googly amplitudes with 2 gluinos. For all these
amplitudes, we got results which are in  agreement with the
previously known results. We note that these googly amplitudes are
simply the ``complex conjugate'' of the corresponding MHV
amplitudes.

The amplitudes with more than two quark-antiquark pairs are no
longer MHV or googly. Nevertheless these amplitudes can be simply
computed by using the extended CSW rules. As an example we will
analyze the purely fermionic amplitudes with 3 quark-antiquark
pairs. There are only 2 different kinds of diagrams and the
amplitude can be written down very simply. We conjecture that this
will be the right result for the amplitude.

This paper is organized as follows. In section 2, we review the
CSW rule for gauge theory without fermions. In section 3, we give
the extended CSW rules for gauge theories with quarks and gluinos.
Here we list all the MHV vertices with fermions. Some general
relations between the number of vertices and the  number of
external gluons are also given. These relations are particularly
useful for analyzing diagrams with fewer number of external gluons
with positive helicity. We will use them to draw the possible
diagrams for the purely fermionic amplitudes with 3
quark-antiquark pairs. In section 4,  we compute the googly
amplitude with one quark-antiquark pair. There is only 1 external
gluon with positive helicity and there can be an arbitrary number
of gluons with negative helicity. In section 5, we compute the
googly amplitude with 2 quark-antiquark pairs. In section 6, we
compute the googly amplitude with 2 gluinos. The computation for
the 4 gluino googly amplitude can also be done and the algebra is
more or less the same as before.  A technical and lengthy proof in
section 4 is relegated to Appendix A.

\section{Review of the CSW approach to perturbative gauge theory}

First let us review the rules for calculating tree level gauge
theory amplitudes as proposed in \cite{Wittena}. Here we follow
the presentation given in \cite{Zhu,Wu} closely.  We will use the
convention that all momenta are outgoing. By MHV (with gluinos
only), we always mean an amplitude with precisely two gluons of
negative helicity. If the two gluons of negative helicity are
labelled as $r,s$ (which may be any integers from $1$ to $n$), the
MHV vertices (or amplitudes) are given as follows:
\begin{equation}
V_n =  {\langle\lambda_r,\lambda_s\rangle^4\over
\prod_{i=1}^n\langle\lambda_i, \lambda_{i+1}\rangle} .
\label{eqone}
\end{equation}
For an on shell (massless) gluon, the momentum in bispinor basis
is given as:
\begin{equation}
p_{a\dot a} = \sigma^\mu_{a\dot a} p_\mu = \lambda_a \tilde{
\lambda}_{\dot a}.
\end{equation}
For an off shell momentum, we can no longer define $\lambda_a$ as
above. The off-shell continuation given in \cite{Wittena} is to
choose an arbitrary spinor $\tilde\eta^{\dot a}$ and then to
define $\lambda_a$ as follows:
\begin{equation}
\lambda_a = p_{a\dot a}\tilde{\eta}^{\dot a}.
\end{equation}
For an on shell momentum $p$, we will use the notation
$\lambda_{pa}$ which is proportional to $\lambda_a$:
\begin{equation}
\lambda_{pa} \equiv  p_{a\dot a} \tilde{\eta}^{\dot a} = \lambda_a
\tilde{\lambda}_{\dot a} \tilde{\eta}^{\dot a} \equiv \lambda_a
\phi_p.
\end{equation}
As demonstrated in \cite{Wittena}, it is consistent to use the
same $\tilde\eta$ for all the off shell lines (or momenta). The
final result is independent of $\tilde{\eta}$.

By using only MHV vertices, one can build a tree diagram by
connecting MHV vertices with propagators.  For the propagator of
momentum $p$, we assign a factor $1/p^2$. Any possible diagram
(involving only MHV vertices) will contribute to the amplitude. As
proved in \cite{Wittena}, a tree level amplitude with $n_-$
external gluons of negative helicity must be obtained from an MHV
tree diagram with $n_--1$ vertices. Another relation was given in
\cite{Zhu},
\begin{equation}
n_+=\sum_{i}\, (i-3)\, n_i+1,\label{eqnp}
\end{equation}
when $n_+$ is the number of the external gluons with positive
helicity, and $n_i$ is the number of the vertices with exactly $i$
line. The other relation stated in the above is:
\begin{equation}
n_-=\sum_{i}n_i+1.
\end{equation}
From eq.~(\ref{eqnp}) we can see that  a tree level amplitude with
$n_+$ external gluons of positive helicity will have no
contribution from any diagram containing an MHV vertex with more
than $n_++2$ lines (not necessarily all internal). For the googly
amplitude we have $n_+=2$. Any contributing diagram will have
exactly one MHV vertex with 4 lines. The rest MHV vertices are all
with 3 lines.

For future use, let us recall 2 formulas presented in \cite{Zhu}
about  the off shell amplitude with $n_+=1$.  When the first
particle with momentum $p_1$ is off shell and has positive
helicity, the amplitude is
\begin{equation}
V_n(1+,2-, \cdots, n-) = {p_1^2 \over \phi_2 \phi_n} \, {1 \over
{[}2,3][3,4]\cdots [n-1,n] }. \label{eq34}
\end{equation}
When it has negative  helicity, the formula is
 \begin{equation}
V_n(1-,2-, \cdots,r+,\cdots,
 n-) = {\phi_r^4 p_1^2 \over \phi_2 \phi_n} \,
{1\over [2,3][3,4]\cdots [n-1,n]} . \label{eq22}
\end{equation}
We stress the fact that the above  off shell amplitudes are
proportional to $p_1^2$ and they  vanish   when $p_1$ is also on
shell ($p_1^2=0$).

\section{The fermionic MHV vertices}

For gauge theory coupled with quark and antiquark, we can
decompose an amplitude into partial amplitudes with definite color
factors \cite{Parkeb}. For simplicity we will assume that all
quarks have different flavors. We will indicate what should be
changed if there are identical quark-antiquark pairs. Also we will
assume the gauge group to be $U(N)$ instead of $SU(N)$. For a
connected diagram with $m$ pair of quark-antiquark, the color
factor is
\begin{equation}
(T^{a_1}   \cdots T^{a_{n_1}})_{i_1\bar{i_2}} ( T^{a_{n_1+1}}
 \cdots T^{a_{n_2}})_{i_2\bar{i_3}} \cdots (
T^{a_{n_{m-1}+1}}   \cdots T^{a_{n}})_{i_m\bar{i_1}} ,
\end{equation}
for a particular ordering of the quark-antiquarks and gluons
\cite{Mangano}. The corresponding partial amplitude is denoted as:
\begin{eqnarray}
& & A(\Lambda_{q_1}^{h_1}, g_1, \cdots, g_{n_1},
\bar{\Lambda}_{{\bar q}_2}^{-h_2}, \Lambda_{q_2}^{h_2}, g_{n_1+1},
\cdots, g_{n_2}, \cdots, \nonumber \\
& & \qquad \qquad \bar{\Lambda}_{{\bar q}_m}^{-h_m} ,
\Lambda_{q_m}^{h_m}, g_{n_{m-1}+1}, \cdots, g_n
\bar{\Lambda}_{{\bar q}_1}^{-h_1} ).
\end{eqnarray}
For amplitudes with connected Feynman diagrams, the
quark-antiquark color indices $(i, \bar i)$ must form a ring of
length exactly $m$. There is no disconnected rings of shorter
length, like in the color factor $(T^{\dots})_{i_1\bar{i_2}}
(T^{\dots})_{i_2\bar{i_1}} $ $(T^{\dots})_{i_3\bar{i_4}}
(T^{\dots})_{i_4\bar{i_3}}$.  This can be proved by induction with
the number of pairs $m$.

The complete amplitude is obtained first by summing over all
possible partitions of $n$ gluons into $m$ parts and their
permutations. Then there is another summation over all possible
permutations between quark-antiquark pairs. If there are identical 
quark-antiquark pairs, we should do one more summation over all
possible permutation between identical antiquarks with a minus
sign if the permutation is odd. Of course, the summations of
different flavor identical quark-antiquark pairs should be done
separately.

\begin{figure}[ht]
    \epsfxsize=80mm%
    \hfill\epsfbox{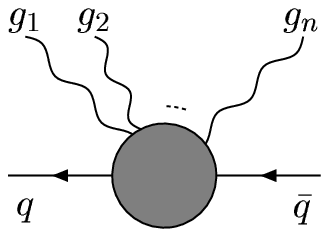}\hfill~\\
    \caption{The graphic representation for the single  pair of
    quark-antiquark partial amplitude. Gluons are emitted from
    one side of the fermion line only.}
     \label{quarkline}
   \end{figure}

For a single quark-antiquark pair the color factor is $ (T^{a_1}
\cdots T^{a_{n}})_{i\bar{i}}$. The partial amplitude is denoted as
$A_{n+2}(\Lambda^h_{q}, g_1, \cdots, g_n, \Lambda^{-h}_{\bar
q})$.\footnote{$h$ denotes the helicity of the quark $q$. The
helicity of the antiquark $\bar q$ is $-h$ by helicity
conservation along the quark line.} It is represented as in
Fig.~\ref{quarkline}. We note that gluon lines are emitted only
from one side of the (connected) quark-antiquark line. We will
stick to this rule also for multi-pair  of quark-antiquark
diagrams.

 \begin{figure}[ht]
    \epsfxsize=100mm%
    \hfill\epsfbox{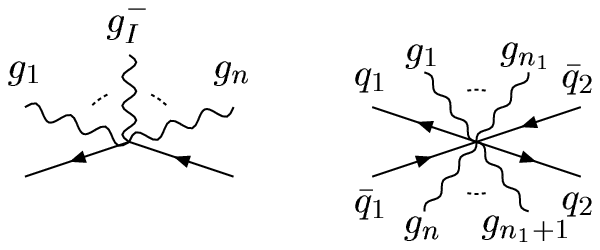}\hfill~\\
    \caption{The 2 MHV vertices with quark-antiquarks.}
         \label{mhvall}
   \end{figure}

There are 2 MHV vertices with quark-antiquarks, one for a single
pair of quark-antiquark and one for two quark-antiquark pairs
which are  shown in Fig.~\ref{mhvall}. There is no MHV vertex for
3 or more pair  of quark-antiquark. All these (non-MHV) amplitudes
should be computed from the above MHV vertices by drawing all
possible (connected) diagrams with only MHV vertices.

The explicit formulas for these MHV amplitudes are given as
follows:
\begin{eqnarray}
& & A(\Lambda_q^+, g_1^+, \cdots, g_I^-, \cdots, g_n^+,
\Lambda_{\bar q}^-) = -{\langle q, I\rangle \langle \bar q, I
\rangle^3 \over \langle q, 1\rangle \langle 1,2 \rangle \cdots
\langle n ,\bar q\rangle
\langle \bar q ,q  \rangle} , \\
& & A(\Lambda_q^-, g_1^+, \cdots, g_I^-, \cdots, g_n^+,
\Lambda_{\bar q}^+) = {\langle q, I\rangle^3 \langle \bar q, I
\rangle \over \langle q, 1\rangle \langle 1,2 \rangle \cdots
\langle n ,\bar q\rangle \langle \bar q ,q  \rangle} ,
\end{eqnarray}
for the single pair of quark-antiquark,  and for 2 quark-antiquark
pairs:
\begin{eqnarray}
 & &\hskip -1cm V(\Lambda_{q_1}^{h_1}, g_1, \cdots, g_{n_1},
 \Lambda_{\bar{ q}_2}^{-h_2}, \Lambda_{q_2}^{h_2}, g_{n_1+1},
 \cdots, g_{n}, \Lambda_{\bar{ q}_1}^{-h_1})  \nonumber\\
 &&= A_0(h_{ 1},h_{ 2})
 {\langle q_1, \bar{q}_2\rangle\over
\langle q_1, 1\rangle \langle  1,  2\rangle \cdots
 \langle n_1, \bar{q_2}\rangle}   \times
{\langle q_2, \bar{q}_1\rangle\over \langle q_2, n_1+1\rangle
  \cdots
 \langle n,\bar{q}_1\rangle}  ,
 \label{eqmhv4f}
\end{eqnarray}
where $A_0(h_{ 1}, h_{ 2})$ is given as follows:
\begin{eqnarray}
A_0(+,+)={\langle \bar{q}_1, \bar{q}_2 \rangle^2 \over \langle
q_1, \bar{q}_1 \rangle \langle q_2, \bar{q}_2\rangle}  , &&
A_0(+,-)=-{\langle \bar{q}_1, q_2 \rangle^2 \over \langle q_1,
\bar{q}_1 \rangle \langle q_2, \bar{q}_2\rangle} , \\
A_0(-,+)=-{\langle q_1, \bar{q}_2 \rangle^2 \over \langle q_1,
\bar{q}_1 \rangle \langle q_2, \bar{q}_2\rangle},  &&
A_0(-,-)={\langle q_1, q_2 \rangle^2 \over \langle q_1, \bar{q}_1
\rangle \langle q_2, \bar{q}_2\rangle} .
\end{eqnarray}
All these MHV amplitudes are given in terms of the ``holomorphic''
spinors of the external (on-shell) momenta. So we can use the same
off shell continuation given in \cite{Wittena} which we recalled
in section 2. By including these fermionic MHV vertices, we can
extend the CSW rule of perturbative gauge theory  to gauge
theories with quarks and antiquarks.   The propagator for both
gluon and fermion (quark or antiquark) internal lines is just
$1/p^2$, as explained in \cite{GeorgiouKhoze}.  The only
peculiarity with fermions is that  helicity is conserved along a
fermion line. Because we assume that all quarks have different
flavor, the flowing of arrows must follow the directions given
exactly in Fig.~\ref{mhvall}.

Let us assume that in an MHV diagram, there are $n_i$ purely
gluonic MHV vertices with $i$-lines, $n_i^{2f}$ single pair of
quark-antiquark MHV vertices with $(i+2)$-lines (counting also the
2 fermion lines, so actually only $i$ gluon lines), and $n_i^{4f}$
2  pairs of quark-antiquark MHV vertices with $(i+4)$-lines
(counting also the 4 fermion lines, so actually only $i$ gluon
lines). For a connected diagram with $2m$ quark-antiquark pairs,
the number of positive helicity gluon $n_+$ and the number of
negative helicity gluon $n_-$ are given as follows:
\begin{eqnarray}
n_- & = & \sum_{i\ge 3} n_i + \sum_{i\ge 1}n_i^{2f} + \sum_{i\ge
0}
n_i^{4f} - (m-1),  \label{numbermm} \\
n_+ & = & \sum_{i\ge 3} (i-3) \, n_i + \sum_{i\ge 1}(i-1)\,
n_i^{2f} + \sum_{i\ge 0} (i+1)\, n_i^{4f} - (m-1) .
\label{numberpp}
\end{eqnarray}

For googly amplitude with $m=1$ and $n_+=1$, we have
\begin{equation}
n_4 + n_2^{2f} + n_0^{4f} = 1,
\end{equation}
and $n_{i>4}=n_{i>2}^{2f}=n_{i>0}^{4f} = 0$. So there is either a
single 4 line gluon MHV vertex, or a 4 line single pair of
quark-antiquark MHV vertex, or a 4 line double pair of
quark-antiquark MHV vertex (which is not possible because we have
only 2 external fermion lines). See Figs.~\ref{onequark1} and
\ref{onequark3}. We will use this result in section 4.

 \begin{figure}[ht]
    \epsfxsize=100mm%
    \hfill\epsfbox{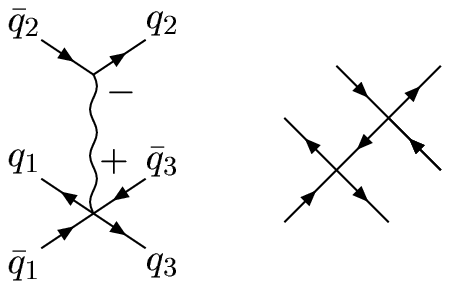}\hfill~\\
    \caption{The 2 different kinds of diagrams  contributing to
    the purely fermionic amplitude with 3 quark-antiquark pairs.}
           \label{6quark}
   \end{figure}

Eqs.~(\ref{numbermm}) and (\ref{numberpp}) are quite powerful for
analyzing the possible diagrams. These relations are particularly
useful for analyzing diagrams with fewer number of external gluons
with positive helicity. For the purely fermionic amplitudes with 3
quark-antiquark pairs, we found that there are  only 2 different
kinds of diagrams  as shown in Fig.~\ref{6quark}. By using the
extended CSW rules, the partial amplitude can be written down very
simply as follows:
\begin{equation}
A^{CSW}_6(\Lambda_{q_1}^-,\Lambda_{\bar
q_2}^+,\Lambda_{q_2}^-,\Lambda_{\bar
q_3}^+,\Lambda_{q_3}^-,\Lambda_{\bar q_1}^+)=\sum_{i=1}^3
A^i+\sum_{i=1}^3 \tilde A^i,
\end{equation}
where
\begin{eqnarray}
A^i&=&-{\langle q_i,(\bar q_i\,q_i)\rangle^3 \, \langle \bar
q_i,(\bar q_i\,q_i)\rangle \over \langle \bar q_i,
q_i\rangle\,\langle q_i,(\bar q_i\,q_i)\rangle\,\langle (\bar
q_i\,q_i), \bar q_i\rangle}\,{1 \over (p_{\bar
q_i}+p_{q_i})^2}\nonumber
\\&& \times
{\langle q_{i+1}, q_{i+2}\rangle^2\over\langle \bar
q_{i+1},q_{i+1}\rangle\, \langle \bar
q_{i+2},q_{i+2}\rangle}\,{\langle q_{i+2}, \bar
q_{i+1}\rangle\over \langle q_{i+2}, (\bar q_i \,  q_i) \rangle
\langle (\bar q_i \, q_i), \bar q_{i+1} \rangle }\nonumber \\
&=&{\langle q_i,(\bar q_i\,q_i)\rangle^2  \over \langle \bar q_i,
q_i\rangle}\,{1 \over (p_{\bar q_i}+p_{q_i})^2}\, {\langle
q_{i+1}, q_{i+2}\rangle^2\over\langle \bar
q_{i+1},q_{i+1}\rangle\, \langle \bar
q_{i+2},q_{i+2}\rangle}\nonumber
\\&& \times{\langle q_{i+2}, \bar
q_{i+1}\rangle\over \langle q_{i+2}, (\bar q_i \,  q_i) \rangle
\langle (\bar q_i \, q_i), \bar q_{i+1} \rangle },
\end{eqnarray}
and
\begin{eqnarray}
\tilde A^i&=&-{\langle q_i, (\bar q_i \, \bar q_{i+1})\rangle^2
\over \langle \bar q_i, q_i\rangle \, \langle \bar q_{i+1}, (\bar
q_i\, q_i) \rangle} \,{1\over (p_{\bar q_i}+p_{q_i}+p_{\bar
q_{i+1}})^2}\\ \nonumber &&\times  {\langle q_{i+1},
q_{i+2}\rangle^2 \over \langle (\bar q_{i+2}\, q_{i+2}),
q_{i+1}\rangle \,\langle \bar q_{i+2}, q_{i+2}\rangle},
\end{eqnarray}
for one set of quark helicities. Here the expression $\langle(ij),
k\rangle$ is defined as $\langle \lambda_{p_i+p_j}, \lambda_k
\rangle$, and the indices are understood mod $3$. We mention that
the six-quark amplitude had been computed before in \cite{Kunszt}.
For other quark helicities and the proof that the above amplitudes
indeed agree with the standard field theory results by using the
Feynman rules, we refer to \cite{Sunwu}.

For  supersymmetric theories, there are also gluinos, the super
partners of the gluons. Because the gluinos are in the same
adjoint representation as the gluons, the color decomposition for
amplitudes with both gluons and gluinos would be the same as for
purely gluon amplitudes.  Compared with gauge theory with
quark-antiquarks, there is no difference between gluino and its
antiparticle. So we should include also diagrams with gluon lines
emitted from both sides of the gluino lines.

The gluino MHV amplitudes can be obtained by using the recursive
relation in \cite{Giele} or by using the supersymmetric Ward
identity \cite{Grisarua,Grisarub, Parkeb} and are recalled here.
The MHV vertex with two gluinos ($r<s$) is:
\begin{eqnarray}
& &V_n(g_1,\cdots,g_t^-,\cdots,\Lambda_r^-,\cdots,\Lambda_s^+,
\cdots,g_n)={\langle t ,r\rangle^3\langle
t,s\rangle\over\prod_{i=1}^n\langle i,i+1\rangle} ,
\label{eqmhvf1}
\end{eqnarray}
where all gluons except one (denoted as $g_t$) have positive
helicity and the gluinos are denoted as $\Lambda_{r,s}$ with their
helicities. As we noted earlier, the helicities along a fermion
line must be conserved. The other case when the positive helicity
gluino is in front of the negative helicity gluino can be obtained
from the above formula by cyclic permutation. One should only note
that there is an extra $-$ sign when we change the order of two
fermions because of  Fermi statistics. So we have:
\begin{eqnarray}
& &V_n(g_1^+,\cdots,g_t^-,\cdots,\Lambda_s^+,\cdots,\Lambda_r^-,
\cdots,g_n^+)=-{\langle t,r\rangle^3 \langle t,s\rangle \over
\prod_{i=1}^n \langle i,i+1\rangle},\label{eqmhvf2}
\end{eqnarray}
for $s<r$ \cite{GeorgiouKhoze}.  From the above two equations, we
see that the position of the negative helicity gluon is
immaterial.

The MHV vertices with $4$ gluinos  is:
\begin{equation}
V_n(g_1,\cdots,\Lambda_p^-,\cdots, \Lambda_q^-, \cdots,
\Lambda_r^+,\cdots,\Lambda_s^+,\cdots,g_n)= -{\langle
r,s\rangle^3\langle p,q \rangle\over \prod_{i=1}^n\langle
i,i+1\rangle}, \label{eq4fermions}
\end{equation}
where all gluons have positive helicity. We note that the above
formula is antisymmetric when we exchange   two gluinos with the
same helicity.

We propose that one can use the above MHV vertices to calculate
all the tree-level (partial) amplitudes by extending the  CSW
approach to   gauge theory with quarks and gluinos. We use the
same off shell continuation given in \cite{Wittena}. In the
following three sections we will test these rules by calculating
all the googly amplitudes with quarks and gluinos. As one can see
from the intermediate steps of our computations, we obtained quite
explicit formulas for some off shell amplitudes. It would be
interesting to explore the connection of these amplitudes with the
non-MHV vertices introduced in \cite{Bernc}.

\section{The googly amplitudes with one quark-anti-quark pair}

In this section we will compute the   googly amplitude with one
quark-antiquark pair which we reported briefly in \cite{Wu}. We
will present here the full details of the computations. The other
cases with 2 quark-antiquark pairs and the supersymmetric case
with gluinos, will be discussed in the next two sections. In this
section we will prove the following formulas for the googly
amplitudes (only the $I$-th gluon has positive helicity):
\begin{eqnarray}
& & A(\Lambda_q^+, g_1^-, \cdots, g_I^+, \cdots, g_n^-,
\Lambda_{\bar q}^-) = {[ q, I]^3 [ \bar q, I ] \over [ q 1] [ 1,2
] \cdots [ n ,\bar q]
[ \bar q ,q  ]} , \\
& & A(\Lambda_q^-, g_1^-, \cdots, g_I^+, \cdots, g_n^-,
\Lambda_{\bar q}^+) =-{ [ q, I] [\bar q, I ]^3 \over  [ q 1] [ 1,2
] \cdots [ n ,\bar q] [ \bar q ,q  ] } . \label{eqer2}
\end{eqnarray}

 \begin{figure}[ht]
    \epsfxsize=80mm%
    \hfill\epsfbox{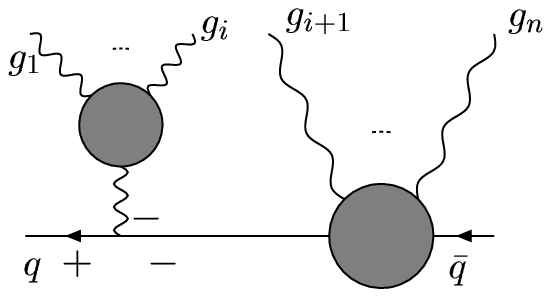}\hfill~\\
    \caption{The decomposition for  the off shell amplitude with one
quark-antiquark pair when the  gluons are all
    with positive helicities   and the quark is off-shell.}
    \label{quark}
   \end{figure}

In order to compute the above googly amplitudes, we first
calculate the amplitudes with one quark-antiquark pair when all
gluons have negative helicity and only one particle is off-shell.
This amplitude would vanish when all particles are on shell. So we
expect that it is proportional to $p^2$ ($p$ is the momentum of
the only off-shell particle). As one case see from
eq.~(\ref{numberpp}), for $n_+=0$ and $m=1$, only $n_3$ and
$n^{2f}_1$ could be non-vanishing. So all the contributing
diagrams are composed with $3$-line MHV vertices only.

To start, let us begin with $n_-=1$. When all the   $3$ particles
are off shell, the MHV amplitude  with one quark-antiquark pair
can be written as:
\begin{eqnarray}
& &A(\Lambda_q^+, g_1^-, \Lambda_{\bar q}^-) ={\langle1,\bar
q\rangle^2\over\langle q, \bar q \rangle} = \langle q,1 \rangle
=\langle 1,\bar q \rangle =\langle \bar q, q \rangle, \label{eq3f} \\
& &A(\Lambda_q^-, g_1^-, \Lambda_{\bar q}^+)=-{\langle1,\bar
q\rangle^2\over\langle q, \bar q \rangle} =-\langle q,1 \rangle
=-\langle 1,\bar q \rangle =-\langle \bar q,  q \rangle .
\end{eqnarray}
The amplitude $A(\Lambda_q^+, g_1^-, \cdots, g_n^-,\Lambda_{\bar
q}^-)$ when $\Lambda_q$ is off shell and has positive helicity, is
given as follows:
\begin{equation}
A(\Lambda_q^+, g_1^-, \cdots, g_n^-,\Lambda_{\bar q}^-)={p_q^2
\over \phi_1} {1\over [1,2][2,3]\cdots[n-1,n][n,\bar
q]}.\label{eqoffshell1}
\end{equation}
As in \cite{Zhu}, we will prove this formula   by mathematical
induction with the number of gluons $n=n_-$.

When $n=1$, from momentum conservation, we have
\begin{equation}
\lambda_q+\lambda_1\phi_1+\lambda_{\bar q} \phi_{\bar q}=0 ,
\end{equation}
and so
\begin{eqnarray}
A(\Lambda_q^+, g_1^-, \Lambda_{\bar q}^-)={\langle 1, {\bar
q}\rangle\over \phi_1}={p_q^2\over \phi_1}{1\over [1,\bar q]}.
\end{eqnarray}
This shows that eq.~(\ref{eqoffshell1}) is true for $n=1$. Now we
assume that it is valid for all $k<n$. We will show that it is
also valid for $k=n$.

In order to compute the amplitude with $n$ gluons we use the
diagram decomposition as shown in Fig.~\ref{quark}. By using the
assumed result for less number of gluons and also
eq.~(\ref{eq34}), we have
\begin{eqnarray}
&&\hskip -2cm A(\Lambda_q^+, g_1^-, \cdots, g_n^-,\Lambda_{\bar
q}^-)= \sum_{i=1}^n {p^2\over \phi_1\phi_i}{1\over [1,2]\cdots
[i-1,i]}\times{1\over p^2}\nonumber\\
&&\times{\langle \lambda_p,\lambda_k\rangle^2\over \langle
\lambda_k, \lambda_1\rangle} \times{1\over k^2} \times{k^2\over
\phi_{i+1}}{1\over [i+1,i+2]\cdots[n-1,n][n,\bar
q]},\label{eqfsum1}
\end{eqnarray}
where
\begin{eqnarray}
p = \sum_{l=1}^i  p_l , \qquad  & &
\lambda_p = \sum_{l=1}^i \lambda_l \phi_l, \\
k = \sum_{l=i+1}^{n+1}  p_l , \qquad & & \lambda_k =
\sum_{l=i+1}^{n+1} \lambda_l \phi_l ,
\end{eqnarray}
and the index $n+1$ (whenever it appears) refers to the antiquark
$\Lambda_{\bar q}$. We note that the degenerate cases for $i=1$
and $i=n$ are also included correctly in the above sum over $i$.

By using eq.~(\ref{eq3f}), we have
\begin{eqnarray}
A(\Lambda_q^+, g_1^-, \cdots, g_n^-,\Lambda_{\bar q}^-)&=& {1\over
\phi_1}{1\over [1,2][2,3]\cdots[n-1,n][n,\bar
q]}\nonumber \\
& &\times
\sum_{i=1}^{n}{[i,i+1]\over\phi_i\phi_{i+1}}\langle\lambda_p,
\lambda_k \rangle. \label{eqerba}
\end{eqnarray}

By using the following identity,
\begin{equation}
\sum_{i=2}^{n-1} {[i, i+1]\over \phi_i \phi_{i+1}} \, \langle
\lambda_{p_2 + \cdots + p_i}, \lambda_{p_{i+1} + \cdots + p_n}
\rangle = (p_2 + p_3 + \cdots + p_n)^2, \label{idb}
\end{equation}
given in \cite{Zhu}, we have
\begin{equation}
\sum_{i=1}^{n}{[i,i+1]\over\phi_i\phi_{i+1}}\langle\lambda_p,
\lambda_k \rangle=(p_1+\cdots+p_n+p_{\bar q})^2=p_q^2.
\end{equation}
By using this result in eq.~(\ref{eqerba}), we have
\begin{equation}
A(\Lambda_q^+, g_1^-, \cdots, g_n^-,\Lambda_{\bar q}^-)=
{p_q^2\over \phi_1}{1\over [1,2][2,3]\cdots[n-1,n][n,\bar q]} ,
\end{equation}
which is the result of eq.~(\ref{eqoffshell1}) for $n+1$ gluons.
This completes the proof of eq.~(\ref{eqoffshell1}).

By using the same method, one can obtain the following result:
\begin{equation}
A(\Lambda_q^-, g_1^-, \cdots, g_n^-,\Lambda_{\bar q}^+)=-{p_q^2
\over \phi_1} {1\over [1,2][2,3]\cdots[n-1,n][n,\bar q]},
\end{equation}
when the quark $\Lambda_q$ is off shell and has negative helicity.

When the anti-quark $\Lambda_{\bar q}$ is off shell, one can
similarly obtain the following results:
\begin{eqnarray}
A(\Lambda_q^+, g_1^-, \cdots, g_n^-,\Lambda_{\bar q}^-)= {p_{\bar
q}^2 \phi_q^2\over \phi_n}{1\over [q,1][1,2][2,3]\cdots[n-1,n]},
\label{eqoffshell2}\\
A(\Lambda_q^-, g_1^-, \cdots, g_n^-,\Lambda_{\bar q}^+)= -{p_{\bar
q}^2 \phi_q^2\over \phi_n}{1\over [q,1][1,2][2,3]\cdots[n-1,n]} .
\end{eqnarray}
When the off shell particle is one of the gluons $g_i$, the
results are:
\begin{eqnarray}
& &\hskip -2cm A(\Lambda_q^+, g_1^-, \cdots, g_n^-,\Lambda_{\bar
q}^-)={p_i^2\phi_q^3\phi_{\bar q} \over
\phi_{i-1}\phi_{i+1}}\nonumber \\
& &\times{1\over [q,1][1,2]\cdots[i-2,i-1][i+1,i+2]\cdots[n,\bar
q][\bar q, q]},
\label{eqoffshell3}\\
& &\hskip -2cm A(\Lambda_q^-, g_1^-, \cdots, g_n^-,\Lambda_{\bar
q}^+)=-{p_i^2\phi_q^3\phi_{\bar q}\over
\phi_{i-1}\phi_{i+1}}\nonumber\\
&& \times{1\over [q,1][1,2]\cdots[i-2,i-1][i+1,i+2]\cdots[n,\bar
q][\bar q, q]}. \label{eqoffshell4}
\end{eqnarray}
Here the index $0$  (whenever it appears) refers to the quark
$\Lambda_q$.

The proof of eq.~(\ref{eqoffshell3}) and eq.~(\ref{eqoffshell4})
is similar. Here we present the proof of eq.~(\ref{eqoffshell3})
only. As we did earlier in the proof of eq.~(\ref{eqoffshell1}),
we will again use   mathematical induction with the number of
gluons $n$.

To start with, it is easy to check that eq.~(\ref{eqoffshell3}) is
true for $n=1$. By assuming that it is true  for all  $k<n$, we
will prove that it is also true for $k=n$.

 \begin{figure}[ht]
    \epsfxsize=70mm%
    \hfill\epsfbox{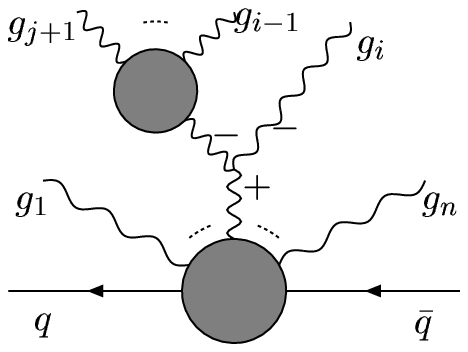}\hfill~\\
    \caption{This diagram is characterized with $g_i$ attaching to a
    3-gluon MHV vertex. $g_{i-1}$'s are disconnected from the fermion
    line by this vertex.}
    \label{gluon1}
   \end{figure}
 \begin{figure}[ht]
    \epsfxsize=70mm%
    \hfill\epsfbox{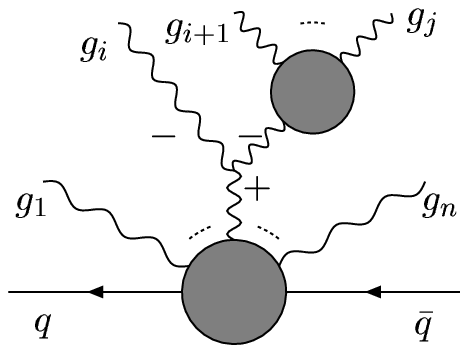}\hfill~\\
    \caption{The same kind diagram as in Fig.~\ref{gluon1}. Here
    it is $g_{i+1}$'s which are disconnected from the fermion
    line by the 3-gluon MHV vertex. }
    \label{gluon2}
   \end{figure}
 \begin{figure}[ht]
    \epsfxsize=90mm%
    \hfill\epsfbox{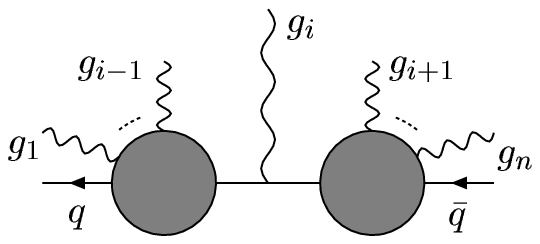}\hfill~\\
    \caption{In this diagram, $g_i$ is directly attached to the
    fermion line.}
    \label{gluon3}
   \end{figure}

Because only $g_i$ is off-shell, one can classify all contributing
diagrams according to the (3-line) MHV vertices attached to $g_i$.
There are 3 kinds of diagrams, shown in Fig.~\ref{gluon1},
Fig.~\ref{gluon2} and Fig.~\ref{gluon3} respectively.  By using
the assumed result for all less multi-particle amplitudes and
eq.~(\ref{eq34}), we can compute the contributions from these
diagrams. The contribution from Fig.~\ref{gluon1}  is
\begin{eqnarray}
 A^1 & = & \sum_{j=0}^{i-2}{\phi_q^3\phi_{\bar q}\over \phi_j
\phi_{i+1}} {1\over
[q,1]\cdots[j-1,j][i+1,i+2]\cdots[n,\bar q][\bar q, q]}\nonumber\\
& &\times{1\over\phi_{j+1}\phi_{i-1}}{1\over [j+1,
j+2]\cdots[i-2,i-1]}\times \langle\lambda_{\tilde
p},\lambda_{\tilde k}\rangle\nonumber\\
&  = & {1\over [q,1][1,2]\cdots[i-2,i-1][i+1,i+2]\cdots[n,\bar
q][\bar q, q]}\nonumber\\
& &\times {\phi_q^3 \phi_{\bar q}\over
\phi_{i-1}\phi_{i+1}}\sum_{j=0}^{i-2}{[j, j+1]\over \phi_j
\phi_{j+1}} \langle\lambda_{\tilde p},\lambda_{\tilde k}\rangle,
\end{eqnarray}
where
\begin{equation} \lambda_{\tilde p}=\sum_{l=0}^j \lambda_l
\phi_l+\sum_{l=i+1}^{n+1}\lambda_l \phi_l, \qquad \lambda_{\tilde
k}=\sum_{l=j+1}^{i-1}\lambda_l \phi_l.
\end{equation}

The contribution from Fig.~\ref{gluon2} can be calculated
similarly as above and the result is:
\begin{eqnarray}
  A^2 & = & \sum_{j=i+1}^{n}{1\over\phi_{i+1}\phi_j}{1\over [i+1,
i+2]\cdots[j-1,j]} \, {\phi_q^3\phi_{\bar q}\over \phi_{i-1}
\phi_{j+1}}
\nonumber\\
& & \times {1\over [q,1]\cdots[i-2, i-1][j+1,j+2]\cdots[n,\bar
q][\bar q, q]}\langle\lambda_{\tilde p},\lambda_{\tilde
k}\rangle\nonumber\\
 &=  & {1\over [q,1][1,2]\cdots[i-2,i-1][i+1,i+2]\cdots[n,\bar
q][\bar q, q]}\nonumber\\
& &\times {\phi_q^3 \phi_{\bar q}\over
\phi_{i-1}\phi_{i+1}}\sum_{j=i+1}^{n}{[j, j+1]\over \phi_j
\phi_{j+1}}\langle\lambda_{\tilde p},\lambda_{\tilde k}\rangle,
\end{eqnarray}
where
\begin{equation}
\lambda_{\tilde p}=\sum_{l=i+1}^{j}\lambda_l \phi_l, \qquad
\lambda_{\tilde k}=\sum_{l=0}^{i-1} \lambda_l
\phi_l+\sum_{l=j+1}^{n+1}\lambda_l \phi_l.
\end{equation}

The last contribution from Fig.~\ref{gluon3} can be calculated by
using eq.~({\ref{eqoffshell1}}) and eq.~(\ref{eqoffshell2}) and we
have:
\begin{eqnarray}
 A^3 & = & {\phi_q^2 \over \phi_{i-1}}{1\over
[q,1]\cdots[i-2,i-1]}\nonumber\\
&&\times{1\over \phi_{i+1}}{1\over [i+1,i+2]\cdots[n+1,\bar q]}
\times \langle\lambda_{\tilde p},
\lambda_{\tilde k}\rangle \nonumber\\
&= & {1\over [q,1][1,2]\cdots[i-2,i-1][i+1,i+2]\cdots[n,\bar
q][\bar q,
q]}\quad \nonumber\\
& &\times {\phi_q^3\phi_{\bar q}\over \phi_{i-1}\phi_{i+1}}{[\bar
q, q] \over \phi_{\bar q}\phi_q}\langle\lambda_{\tilde
p},\lambda_{\tilde k}\rangle,
\end{eqnarray}
where
\begin{equation}
\lambda_{\tilde p}=\sum_{l=i+1}^{n+1}\lambda_l \phi_l, \qquad
\lambda_{\tilde k}=\sum_{l=0}^{i-1} \lambda_l \phi_l.
\end{equation}

By combing all these three contributions together and using the
following result:
\begin{equation}
\sum_{j=0}^{i-2}{[j, j+1]\over \phi_j
\phi_{j+1}}\langle\lambda_{\tilde p},\lambda_{\tilde
k}\rangle+{[\bar q, q] \over \phi_{\bar
q}\phi_q}\langle\lambda_{\tilde p},\lambda_{\tilde
k}\rangle+\sum_{j=i+1}^{n}{[j, j+1]\over \phi_j
\phi_{j+1}}\langle\lambda_{\tilde p},\lambda_{\tilde
k}\rangle=p_i^2.
\end{equation}
from eq.~(\ref{idb}), we have
\begin{eqnarray}
& & \hskip -1cm A(\Lambda_q^+, g_1^-, \cdots, g_n^-,\Lambda_{\bar
q}^-) = \sum_{s=1}^3A^s\, \nonumber\\
 & & = {\phi_q^3 \phi_{\bar q}\over
\phi_{i-1}\phi_{i+1}} \,  {p_i^2\over
[q,1][1,2]\cdots[i-2,i-1][i+1,i+2]\cdots[n,\bar q][\bar q, q]} .
\end{eqnarray}
This ends the proof of eq.~(\ref{eqoffshell3}).

 \begin{figure}[ht]
    \epsfxsize=90mm%
    \hfill\epsfbox{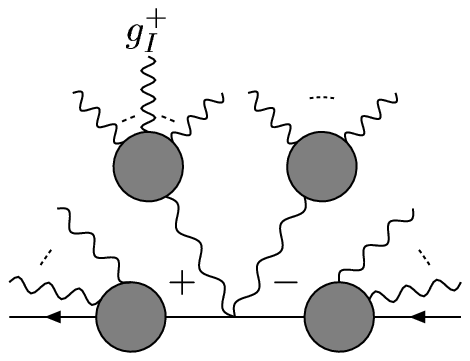}\hfill~\\
    \caption{One kind of diagrams from the 4-particle MHV vertex with
    one quark-antiquark pair.}
    \label{onequark1}
   \end{figure}

Now we begin to compute the googly amplitudes with one
quark-anti-quark pair. As we noted in section 3, for googly
amplitude, there is only one 4-particle MHV vertex in any
contributing diagram. The rest vertices are 3-particle MHV
vertices. So we can classify all contributing diagrams according
to this 4-particle MHV vertex. Form the 4-particle MHV vertex with
one quark-antiquark pair, there are 2 kinds of diagrams. One kind
of diagrams is shown Fig.~\ref{onequark1} where the only  positive
helicity gluon $g_I$ is in the left blob connected to the
4-particle vertex. The other kind of diagrams (which is not shown)
can be obtained from Fig.~\ref{onequark1} by interchanging the two
gluon lines connected to the 4-particle MHV  vertex. The
contributions from these two kinds of diagrams are:
\begin{eqnarray}
A^1&=&{\phi_q^2 \phi_I^4 \over [q,1][1,2]\cdots [n-1,n][n, \bar
q]}\sum_{i=0}^{I-1}\sum_{j=I}^{n-1}\sum_{k=j+1}^n {[i,i+1]\over
\phi_i \phi_{i+1}} \,{[j,j+1]\over \phi_j \phi_{j+1}}
\,\nonumber \\
& & \times{[k,k+1]\over \phi_k \phi_{k+1}}  \,  {-\lan V_1,
V_3\ran \lan V_4, V_3 \ran^3\over \lan V_1, V_2 \ran \lan V_2,
V_3\ran \lan V_3, V_4\ran \lan V_4, V_1\ran},
\end{eqnarray}
and
\begin{eqnarray}
A^2&=&{\phi_q^2 \phi_I^4 \over [q,1][1,2]\cdots [n-1,n][n, \bar
q]}\sum_{i=0}^{I-2}\sum_{j=i+1}^{I-1}\sum_{k=I}^n {[i,i+1]\over
\phi_i \phi_{i+1}} \,
{[j,j+1]\over \phi_j \phi_{j+1}} \,\nonumber \\
& & \times {[k,k+1]\over \phi_k \phi_{k+1}}\,  {-\lan V_1, V_2\ran
\lan V_4, V_2 \ran^3\over \lan V_1, V_2 \ran \lan V_2, V_3\ran
\lan V_3, V_4 \ran \lan V_4, V_1\ran},
\end{eqnarray}
respectively.  Here
\begin{eqnarray}
V_1=\sum_{s=0}^i \lambda_s \phi_s, & & \quad V_2=\sum_{s=i+1}^j
\lambda_s \phi_s,\label{eqfv1}\\
 V_3=\sum_{s=j+1}^k \lambda_s \phi_s, & & \quad
V_4=\sum_{s=k+1}^{n+1} \lambda_s \phi_s.\label{eqfv2}
\end{eqnarray}
The sum of $A^1$ and $A^2$ can be written as follows:
\begin{eqnarray}
A^1+A^2&=&{\phi_q^2 \phi_I^4 \over [q,1][1,2]\cdots [n-1,n][n,
\bar q]}\sum_{i=0}^{I-1}\,\sum_{k={\rm max}\{I,
i+2\}}^n\,\sum_{j=i+1}^{k-1} {[i,i+1]\over \phi_i \phi_{i+1}}
\,\nonumber \\
& & \times {[j,j+1]\over \phi_j \phi_{j+1}}
 \,{[k,k+1]\over \phi_k \phi_{k+1}} \, {-\lan V_1, V_p\ran
\lan V_4, V_p \ran^3\over \lan V_1, V_2 \ran \lan V_2, V_3\ran\lan
V_3, V_4 \ran \lan V_4, V_1\ran},\label{eqfg1}
\end{eqnarray}
where $p$ in eq.~(\ref{eqfg1}) is the index ($p=2,3$) which
doesn't include $\lambda_I \phi_I$ as defined in
eqs.~(\ref{eqfv1}) and (\ref{eqfv2}).

   \begin{figure}[ht]
    \epsfxsize=80mm%
    \hfill\epsfbox{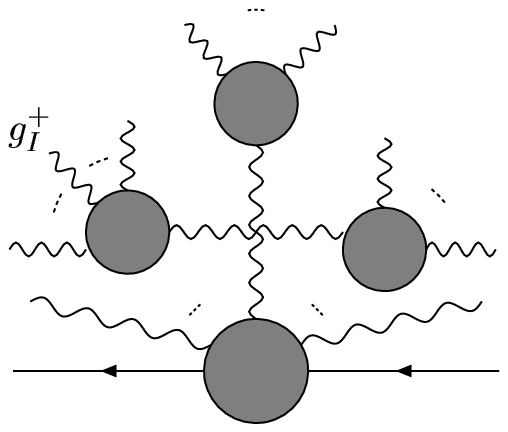}\hfill~\\
    \caption{One kind of diagrams from the 4-gluon MHV vertex. Here the
    only positive helicity gluon $g_I$ is in the far left blob.}
    \label{onequark3}
   \end{figure}

From the 4-gluon MHV vertex,  there are 3 kinds of diagrams
depending on the position   of the only positive helicity gluon
$g_I$. One kind of diagrams is shown in Fig.~\ref{onequark3} where
the gluon $g_I$ is in the left blob connected to the 4-gluon MHV
vertex. The other 2 kinds of diagrams (which are not shown here)
are the same diagrams but with the gluon $g_I$ staying middle blob
or in the   right blob. The contributions from these diagrams are:
\begin{eqnarray}
\sum_{t=3}^5 A^t&=&{\phi_q^3 \phi_{\bar q} \phi_I^4 \over
[q,1][1,2]\cdots [n-1,n][n, \bar q][\bar q,
q]}\sum_{i=0}^{I-1}\,\sum_{l={\rm max}\{I,
i+3\}}^n\,\sum_{j=i+1}^{l-2}\sum_{k=j+1}^{l-1} {[i,i+1]\over
\phi_i \phi_{i+1}} \,
\nonumber \\
& & \times {[j,j+1]\over \phi_j \phi_{j+1}} \,{[k,k+1]\over \phi_k
\phi_{k+1}}{[l,l+1]\over \phi_l \phi_{l+1}}\, {\lan \tilde
V_r,\tilde V_s \ran^4\over \lan \tilde V_1, \tilde V_2 \ran \lan
\tilde V_2, \tilde V_3\ran \lan \tilde V_3,\tilde V_4 \ran \lan
\tilde V_4,\tilde V_1\ran},\label{eqfg2}
\end{eqnarray}
where
\begin{eqnarray}
&& \tilde V_1=\sum_{s=0}^i \lambda_s \phi_s+\sum_{s=l+1}^{n+1}
\lambda_s \phi_s,  \quad \tilde V_2=\sum_{s=i+1}^j \lambda_s
\phi_s, \label{eqftv1}\\
&& \tilde V_3=\sum_{s=j+1}^k \lambda_s \phi_s,  \qquad \quad
\tilde V_4=\sum_{s=k+1}^l \lambda_s \phi_s,\label{eqftv2}
\end{eqnarray}
and  $r$ and $s$ in eq.~(\ref{eqfg2}) are the two indexes
($r=2,3$, $s=3,4$, $r\ne s$) which satisfy that neither $\tilde
V_r$ nor $\tilde V_s$ includes $\lambda_I\phi_I$ as defined in
eqs.~(\ref{eqftv1}) and (\ref{eqftv2}).

By combining all these contributions, the googly amplitude is
\begin{equation}A(\Lambda_q^+, g_1^-,\cdots, g_I^+, \cdots, g_n^-,
 \Lambda_{\bar q})=\sum_{i=1}^5 A^i.
\end{equation}

In Appendix A, we will prove the following identity:
\begin{eqnarray}
& & \hskip -.5cm {[\bar q, q]\over \phi_{\bar
q}\phi_q}\sum_{i=0}^{I-1}\sum_{k={\rm max}\{I, i+2\}}^n
\,\sum_{j=i+1}^{k-1} {[i,i+1]\over \phi_i \phi_{i+1}} \,
{[j,j+1]\over \phi_j \phi_{j+1}}
 \,{[k,k+1]\over \phi_k \phi_{k+1}} \nonumber\\
 & &\times {-\lan V_1, V_p\ran
\lan V_4, V_p \ran^3\over \lan V_1, V_2 \ran \lan V_2, V_3\ran
\lan V_3, V_4 \ran\lan V_4, V_1\ran}\nonumber \\
& & +\sum_{i=1}^{I-1}\,\sum_{l={\rm max}\{I,
i+3\}}^n\,\sum_{j=i+1}^{l-2}\sum_{k=j+1}^{l-1} {[i,i+1]\over
\phi_i \phi_{i+1}} \,{[j,j+1]\over \phi_j \phi_{j+1}}
\,{[k,k+1]\over \phi_k \phi_{k+1}}
\nonumber \\
& & \times {[l,l+1]\over \phi_l \phi_{l+1}}\, {\lan \tilde
V_r,\tilde V_s \ran^4\over \lan \tilde V_1, \tilde V_2 \ran \lan
\tilde V_2, \tilde V_3\ran \lan \tilde V_3,\tilde V_4 \ran \lan
\tilde V_4,\tilde V_1\ran}=  {[q, I]^3[\bar q, I] \over
\phi_q^3\phi_{\bar q}}. \label{eqidf}
\end{eqnarray}
By using this identity, we have:
\begin{equation}
A(\Lambda_q^+, g_1^+, \cdots, g_I^-, \cdots, g_n^+, \Lambda_{\bar
q}^-) = {[ q, I]^3 [ \bar q, I ] \over [ q 1] [ 1,2] \cdots [n
,\bar q] [ \bar q ,q ]},
\end{equation}
which  is the expected result for the googly amplitude and it is
in agreement with the result obtained by other methods
\cite{Parkeb}.

The calculation for $A(\Lambda_q^-, g_1^+, \cdots, g_I^-, \cdots,
g_n^+, \Lambda_{\bar q}^+)$ is similar and the result is
eq.~(\ref{eqer2}).

\section{The 4 quark-anti-quark googly amplitude}

In this section we will compute the googly amplitude with two
quark-antiquark pairs. Here all the gluons have negative helicity.
The googly amplitudes we want to reproduce are given as follows:
\begin{eqnarray}
 & & \hskip -1.5cm A(\Lambda_{q_1}^{h_1},
 l_1, \cdots, l_k,  \Lambda_{\bar{q}_2}^{-h_2},\Lambda_{q_2}^{h_2},
  m_1, \cdots,
 m_{n-k}, \Lambda_{\bar{q}_1}^{-h_1} )\nonumber \\
 &  = & A^{\prime}_0(h_{ 1},h_{ 2})   {[ q_1, \bar{q}_2]\over
[ q_1, l_1] [ l_1, l_2] \cdots
 [l_k, \bar{q_2}]}
{[ q_2, \bar{q}_1]\over [ q_2, m_1] [ m_1, m_2] \cdots [
m_{n-k},\bar{q}_1]}   ,
\end{eqnarray}
where
\begin{eqnarray}
A^{\prime}_0(+,+)={[q_1, q_2]^2 \over [ q_1, \bar{q}_1 ][ q_2,
\bar{q}_2]},  && A^{\prime}_0(+,-)=-{[ q_1, \bar{q}_2 ]^2 \over [
q_1,
\bar{q}_1 ][ q_2, \bar{q}_2]}, \\
A^{\prime}_0(-,+)=-{[\bar{q}_1, q_2 ]^2 \over [ q_1, \bar{q}_1 ] [
q_2, \bar{q}_2]},  && A^{\prime}_0(-,-)={[ \bar{q}_1, \bar{q}_2
 ]^2 \over [ q_1, \bar{q}_1 ] [ q_2,
\bar{q}_2]}.
\end{eqnarray}

As discussed in  section 3, in all of the diagrams which
contribute to the googly amplitudes there is just one MHV vertices
with $4$ lines. All other vertices are with $3$ lines as in the
case for the amplitudes with gluons only \cite{Zhu,Wu}. So we can
classify all contributing diagrams by using this unique 4-particle
MHV vertex.

 \begin{figure}[ht]
    \epsfxsize=90mm%
    \hfill\epsfbox{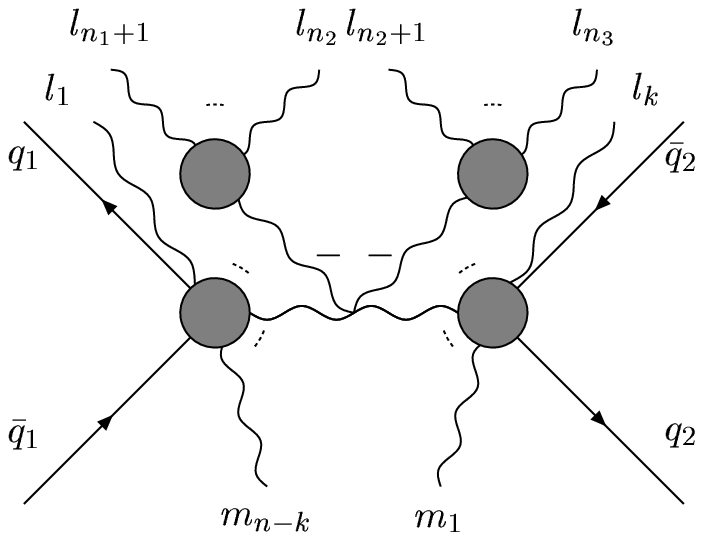}\hfill~\\
    \caption{One kind of diagrams from the 4-gluon MHV vertex. The two
    gluons connected to  the fermion lines have positive helicity. This
    gives contribution $A^1$. Two
    more kinds of the same type of diagrams are obtained by permutating
    4 internal gluon lines which give contributions $A^{2,3}$.}
     \label{twoquark1}
   \end{figure}

First we have one kind of diagrams from the 4-gluon MHV vertex as
shown in Fig.~\ref{twoquark1}.  Here the two gluons connected to
the fermion lines have positive helicity. This contribution is:
\begin{eqnarray}
&&\hskip -1cm A^1=-F_0 \sum_{n_1=0}^{k-2} \sum_{n_2=n_1+1}^{k-1}
\sum_{n_3=n_2+1}^k \sum_{n_4=0}^{n-k}
{\tilde F}^1(n_1,\cdots,n_4) \nonumber\\
&&\times {\lan V^1_2, V^1_3 \ran^3\over \lan V^1_1, V^1_2 \ran
\lan V^1_3, V^1_4\ran \lan V^1_4, V^1_1\ran}, \label{eqonesix}
\end{eqnarray}
where the index $l_0$ ($ l_{k+1}, m_0, m_{n-k+1}$)   refers to
$q_1$ ($\bar{q}_2, q_2, \bar{q}_1$) and $F_0$ is defined as
\begin{equation}
F_0=\phi_{q_1}^3\phi_{\bar{q}_1}\phi_{q_2}^3\phi_{\bar{q}_2}{1
\over [ \bar{q}_1,q_1]}{1 \over [ \bar{q}_2,q_2]}\prod_{i=0}^{k}\,
{1\over [l_i, l_{i+1}]}\, \prod_{j=0}^{n-k}\, {1\over [m_j,
m_{j+1}]}\, .
\end{equation}
The other quantities appearing in eq.~(\ref{eqonesix}) are defined
as follows:
\begin{equation}
{\tilde F}^1(n_1,\cdots,n_4)=\prod_{i=1}^3 {[l_{n_i}, l_{n_i+1}]
\over \phi_{l_{n_i}}\phi_{l_{n_i}+1}}  {[m_{n_4}, m_{n_4+1}] \over
\phi_{m_{n_4}}\phi_{m_{n_4}+1}},
\end{equation}
and
\begin{eqnarray}
&&V^1_1=\sum_{s=1}^{n_1}\lambda_{l_s}\phi_{l_s}+
\sum_{s=n_4+1}^{n-k}\lambda_{m_s}\phi_{m_s} +\lambda_{q_1}
\phi_{q_1}+\lambda_{\bar{q}_1}\phi_{\bar{q}_1},\\
&&V^1_2=\sum_{s=n_1+1}^{n_2}\lambda_{l_s}\phi_{l_s}, \\
&&V^1_3=\sum_{s=n_2+1}^{n_3}\lambda_{l_s}\phi_{l_s}, \\
&&V^1_4=\sum_{s=n_3+1}^k\lambda_{l_s}\phi_{l_s}+
\sum_{s=1}^{n_4}\lambda_{m_s}\phi_{m_s} +\lambda_{q_2}
\phi_{q_2}+\lambda_{\bar{q}_2}\phi_{\bar{q}_2}.
\end{eqnarray}
The other 2 similar contributions are:
\begin{eqnarray}
  A^2 & = & -F_0 \sum_{n_1=0}^{k-1} \sum_{n_2=n_1+1}^{k}
\sum_{n_3=0}^{n-k-1} \sum_{n_4=n_3+1}^{n-k}
{\tilde F}^2(n_1,\cdots,n_4) \nonumber\\
&&\times {\lan V^2_2, V^2_4 \ran^4\over \lan V^2_1, V^2_2 \ran
\lan V^2_2, V^2_3 \ran \lan V^2_3, V^2_4\ran \lan V^2_4,
V^2_1\ran}, \\
  A^3 & = & -F_0 \sum_{n_1=0}^{k} \sum_{n_2=0}^{n-k-2}
\sum_{n_3=n_2+1}^{n-k-1} \sum_{n_4=n_3+1}^{n-k}
{\tilde F}^3(n_1,\cdots,n_4) \nonumber\\
&&\times {\lan V^3_3, V^3_4 \ran^4\over \lan V^3_1, V^3_2 \ran
\lan V^3_2, V^3_3 \ran \lan V^3_3, V^3_4\ran \lan V^3_4,
V^3_1\ran},
\end{eqnarray}
where
\begin{eqnarray}
{\tilde F}^2(n_1,\cdots,n_4) & = & \prod_{i=1}^2 {[l_{n_i},
l_{n_i+1}] \over \phi_{l_{n_i}}\phi_{l_{n_i}+1}}
\prod_{j=3}^4{[m_{n_j}, m_{n_j+1}] \over
\phi_{m_{n_j}}\phi_{m_{n_j}+1}}, \\
{\tilde F}^3(n_1,\cdots,n_4) & = & {[l_{n_1}, l_{n_1+1}] \over
\phi_{l_{n_1}}\phi_{l_{n_1}+1}}  \prod_{j=2}^4{[m_{n_j},
m_{n_j+1}] \over \phi_{m_{n_j}}\phi_{m_{n_j}+1}},
\end{eqnarray}
and
\begin{eqnarray}
&&V^2_1=\sum_{s=1}^{n_1}\lambda_{l_s}\phi_{l_s}+
\sum_{s=n_4+1}^{n-k}\lambda_{m_s}\phi_{m_s} +\lambda_{q_1}
\phi_{q_1}+\lambda_{\bar{q}_1}\phi_{\bar{q}_1},\\
&&V^2_2=\sum_{s=n_1+1}^{n_2}\lambda_{l_s}\phi_{l_s}, \\
&&V^2_3=\sum_{s=n_2+1}^{k}\lambda_{l_s}\phi_{l_s}+
\sum_{s=1}^{n_3}\lambda_{m_s}\phi_{m_s}+\lambda_{q_2}
\phi_{q_2}+\lambda_{\bar{q}_2}\phi_{\bar{q}_2}, \\
&&V^2_4=\sum_{s=n_3+1}^{n_4}\lambda_{m_s}\phi_{m_s}, \\
&&V^3_1=\sum_{s=1}^{n_1}\lambda_{l_s}\phi_{l_s}+
\sum_{s=n_4+1}^{n-k}\lambda_{m_s}\phi_{m_s} +\lambda_{q_1}
\phi_{q_1}+\lambda_{\bar{q}_1}\phi_{\bar{q}_1},\\
&&V^3_2=\sum_{s=n_1+1}^{k}\lambda_{l_s}\phi_{l_s}+
\sum_{s=1}^{n_2}\lambda_{m_s}\phi_{m_s}+\lambda_{q_2}
\phi_{q_2}+\lambda_{\bar{q}_2}\phi_{\bar{q}_2}, \\
&&V^3_3=\sum_{s=n_2+1}^{n_3}\lambda_{l_s}\phi_{l_s}, \\
&&V^3_4=\sum_{s=n_3+1}^{n-k}\lambda_{l_s}\phi_{l_s} .
\end{eqnarray}

\begin{figure}[ht]
    \epsfxsize=90mm%
    \hfill\epsfbox{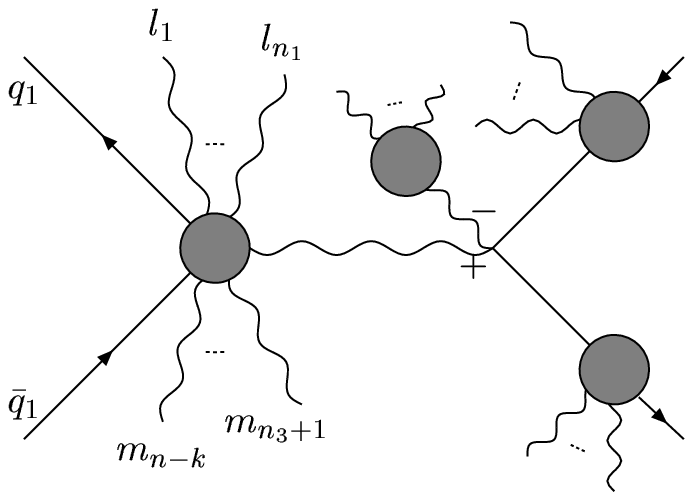}\hfill~\\
    \caption{One kind of diagrams from the 4-particle MHV vertex with
    one quark-antiquark pair. This gives contribution $A^4$.
    There are 3 more kinds of the same type
    of diagrams which give contributions $A^{5,6,7}$. }
    \label{twoquark4}
   \end{figure}

The second group of contributions are from the diagrams with the
4-particle MHV vertex having one quark-antiquark pair. An example
is shown in Fig.~\ref{twoquark4}. Its contribution $A^4$ is given
as follows:
\begin{equation}
  A^4=F_0 \sum_{n_1=0}^{k-1} \sum_{n_2=n_1+1}^{k}
\sum_{n_3=0}^{n-k} {\tilde F}^4(n_1,n_2,n_3) \,
 {\lan V^4_2, V^4_3 \ran^2\lan V^4_2, V^4_4\ran\over \lan
V^4_1, V^4_2 \ran  \lan V^4_3, V^4_4\ran \lan V^4_4, V^4_1\ran},
\end{equation}
where
\begin{equation}
{\tilde F}^4(n_1,n_2,n_3)=\prod_{i=1}^2 {[l_{n_i}, l_{n_i+1}]
\over \phi_{l_{n_i}}\phi_{l_{n_i}+1}}  {[m_{n_3}, m_{n_3+1}] \over
\phi_{m_{n_3}}\phi_{m_{n_3}+1}} {[q_2, \bar{q}_2] \over
\phi_{q_2}\phi_{\bar{q}_2}},
\end{equation}
and
\begin{eqnarray}
&&V^4_1=\sum_{s=1}^{n_1}\lambda_{l_s}\phi_{l_s}+
\sum_{s=n_3+1}^{n-k}\lambda_{m_s}\phi_{m_s} +\lambda_{q_1}
\phi_{q_1}+\lambda_{\bar{q}_1}\phi_{\bar{q}_1},\\
&&V^4_2=\sum_{s=n_1+1}^{n_2}\lambda_{l_s}\phi_{l_s}, \\
&&V^4_3=\sum_{s=n_2+1}^{k}\lambda_{l_s}\phi_{l_s}+\lambda_{\bar{q}_2}\phi_{\bar{q}_2}, \\
&&V^4_4=\sum_{s=1}^{n_3}\lambda_{m_s}\phi_{m_s} +\lambda_{q_2}
\phi_{q_2} .
\end{eqnarray}
There are 3 more kinds of the same type of diagrams which give
contributions $A^{5,6,7}$:
\begin{eqnarray}
  A^5 & = & F_0 \sum_{n_1=0}^{k} \sum_{n_2=0}^{n-k-1}
\sum_{n_3=n_2+1}^{n-k} {\tilde F}^5(n_1,n_2,n_3) \, {\lan V^5_2,
V^5_4 \ran^3\over \lan V^5_1, V^5_2 \ran
\lan V^5_2, V^5_3\ran \lan V^5_4, V^5_1\ran}, \\
  A^6 & = & F_0 \sum_{n_1=0}^{k-1}
\sum_{n_2=n_1+1}^{k} \sum_{n_3=0}^{n-k} {\tilde F}^6(n_1,n_2,n_3)
\,  {-\lan V^6_2, V^6_4 \ran^3\over \lan V^6_2, V^6_3 \ran
\lan V^6_3, V^6_4\ran \lan V^6_4, V^6_1\ran}, \\
  A^7 & = & F_0 \sum_{n_1=0}^{k} \sum_{n_2=0}^{n-k-1}
\sum_{n_3=n_2+1}^{n-k} {\tilde F}^7(n_1,n_2,n_3)  \,  {-\lan
V^7_1, V^7_3 \ran \lan V^7_3, V^7_4\ran^2\over \lan V^7_1,
V^7_2\ran \lan V^7_2, V^7_3 \ran  \lan V^7_4, V^7_1\ran},
\end{eqnarray}
where
\begin{eqnarray}
{\tilde F}^5(n_1,n_2,n_3) & =& {[l_{n_1}, l_{n_1+1}] \over
\phi_{l_{n_1}}\phi_{l_{n_1}+1}} \prod_{j=2}^3 {[m_{n_j},
m_{n_j+1}] \over \phi_{m_{n_j}}\phi_{m_{n_j}+1}} {[q_2, \bar{q}_2]
\over \phi_{q_2}\phi_{\bar{q}_2}}, \\
{\tilde F}^6(n_1,n_2,n_3) & =& \prod_{i=1}^2 {[l_{n_i}, l_{n_i+1}]
\over \phi_{l_{n_i}}\phi_{l_{n_i}+1}}  {[m_{n_3}, m_{n_3+1}] \over
\phi_{m_{n_3}}\phi_{m_{n_3}+1}} {[ \bar{q}_1, q_1] \over
\phi_{\bar{q}_1} \phi_{q_1}}, \\
{\tilde F}^7(n_1,n_2,n_3) & =& {[l_{n_1}, l_{n_1+1}] \over
\phi_{l_{n_1}}\phi_{l_{n_1}+1}} \prod_{j=2}^3 {[m_{n_j},
m_{n_j+1}] \over \phi_{m_{n_j}}\phi_{m_{n_j}+1}} {[ \bar{q}_1,
q_1] \over \phi_{\bar{q}_1} \phi_{q_1}},
\end{eqnarray}
and
\begin{eqnarray}
&&V^5_1=\sum_{s=1}^{n_1}\lambda_{l_s}\phi_{l_s}+
\sum_{s=n_3+1}^{n-k}\lambda_{m_s}\phi_{m_s} +\lambda_{q_1}
\phi_{q_1}+\lambda_{\bar{q}_1}\phi_{\bar{q}_1},\\
&&V^5_2=\sum_{s=n_1+1}^{k}\lambda_{l_s}\phi_{l_s}+\lambda_{\bar{q}_2}\phi_{\bar{q}_2}, \\
&&V^5_3=\sum_{s=1}^{n_2}\lambda_{m_s}\phi_{l_s}+\lambda_{q_2}
\phi_{q_2}, \\
&&V^5_4=\sum_{s=n_2+1}^{n_3}\lambda_{m_s}\phi_{m_s}, \\
&&V^6_1=\sum_{s=1}^{n_1}\lambda_{l_s}\phi_{l_s}+\lambda_{q_1}
\phi_{q_1},\\
&&V^6_2=\sum_{s=n_1+1}^{n_2}\lambda_{l_s}\phi_{l_s}, \\
&&V^6_3=\sum_{s=n_2+1}^{k}\lambda_{l_s}\phi_{l_s}+\sum_{s=1}^{n_3}\lambda_{m_s}\phi_{m_s}
+\lambda_{q_2}
\phi_{q_2}+\lambda_{\bar{q}_2}\phi_{\bar{q}_2}, \\
&&V^6_4=\sum_{s=n_3+1}^{n-k}\lambda_{m_s}\phi_{m_s}+\lambda_{\bar{q}_1}\phi_{\bar{q}_1},
\\
&&V^7_1=\sum_{s=1}^{n_1}\lambda_{l_s}\phi_{l_s}+\lambda_{q_1}
\phi_{q_1},\\
&&V^7_2=\sum_{s=n_1+1}^{k}\lambda_{l_s}\phi_{l_s}+\sum_{s=1}^{n_2}\lambda_{m_s}\phi_{m_s}
+\lambda_{q_2}
\phi_{q_2}+\lambda_{\bar{q}_2}\phi_{\bar{q}_2}, \\
&&V^7_3=\sum_{s=n_2+1}^{n_3}\lambda_{m_s}\phi_{m_s}, \\
&&V^7_4=\sum_{s=n_3+1}^{n-k}\lambda_{m_s}\phi_{m_s}+
\lambda_{\bar{q}_1}\phi_{\bar{q}_1} .
\end{eqnarray}

 \begin{figure}[ht]
    \epsfxsize=80mm%
    \hfill\epsfbox{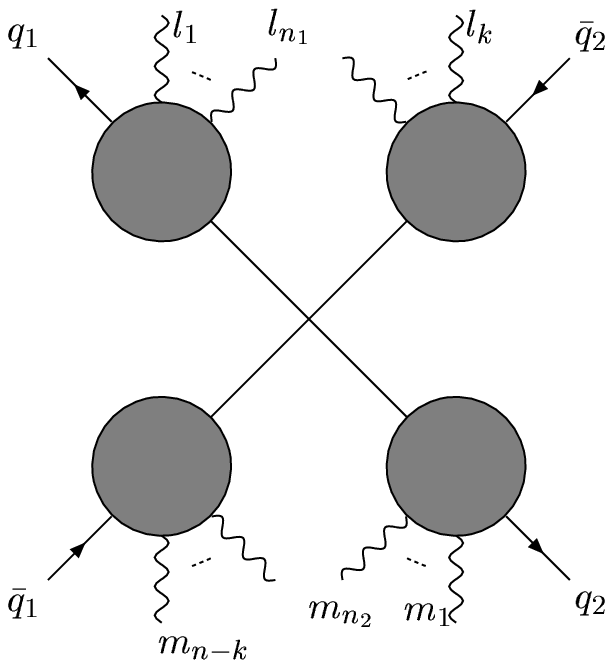}\hfill~\\
    \caption{The only kind of contributing diagrams from
    the 4-fermion MHV vertex. This gives the last contribution
    $A^8$.}
        \label{twoquark8}
   \end{figure}

The last contribution comes from the only kind of contributing
diagrams from the 4-fermion MHV vertex as shown in
Fig.~\ref{twoquark8}. This contribution is:
\begin{equation}
  A^8=F_0 \sum_{n_1=0}^{k} \sum_{n_2=0}^{n-k}
{\tilde F}^8(n_1,n_2) \,  {\lan V^8_2, V^8_4\ran^2\over \lan
V^8_1, V^8_2\ran \lan V^8_3, V^8_4\ran},
\end{equation}
where
\begin{equation}
{\tilde F}^8(n_1,n_2)={[l_{n_1}, l_{n_1+1}] \over
\phi_{l_{n_1}}\phi_{l_{n_1}+1}} {[m_{n_2}, m_{n_2+1}] \over
\phi_{m_{n_2}}\phi_{m_{n_2}+1}} {[ \bar{q}_1, q_1] \over
\phi_{\bar{q}_1} \phi_{q_1}} {[q_2, \bar{q}_2] \over
\phi_{q_2}\phi_{\bar{q}_2}},
\end{equation}
and
\begin{eqnarray}
&&V^8_1=\sum_{s=1}^{n_1}\lambda_{l_s}\phi_{l_s}+\lambda_{q_1}
\phi_{q_1},\\
&&V^8_2=\sum_{s=n_1+1}^{k}\lambda_{l_s}\phi_{l_s}
+\lambda_{\bar{q}_2}\phi_{\bar{q}_2}, \\
&&V^8_3=\sum_{s=1}^{n_2}\lambda_{m_s}\phi_{m_s}+\lambda_{q_2}
\phi_{q_2}, \\
&&V^8_4=\sum_{s=n_2+1}^{n-k}\lambda_{m_s}
\phi_{m_s}+\lambda_{\bar{q}_1}\phi_{\bar{q}_1} .
\end{eqnarray}

By summing over all these contributions and using the same method
for proving identities as in Appendix A,  we have:
\begin{eqnarray}
\sum_{i=1}^8 A^i&=&{[q_1, q_2]^2 \over [ q_1, \bar{q}_1 ][ q_2,
\bar{q}_2]}{[ q_1, \bar{q}_2]\over [ q_1, l_1] [ l_1, l_2] \cdots
 [l_k, \bar{q_2}]}\nonumber \\
 & & \times
{[ q_2, \bar{q}_1]\over [ q_2, m_1] [ m_1, m_2] \cdots [
m_{n-k},\bar{q}_1]} .
\end{eqnarray}
This is the correct result for the googly amplitude with two
quark-antiquark pairs and $n$ gluons.

\section{The gluino googly amplitudes}

In this last section we will compute the googly amplitudes with
gluinos. As in the case without fermions \cite{Zhu}, we first
calculate the off shell amplitudes with all gluons having negative
helicity. From eq.~(\ref{numberpp}) with $n_+=0$ and $m=1$,  only
$n_3$ and $n_1^{2f}$ could be non-vanishing. So all contributing
diagrams compose of 3-line MHV vertices. Again we will find that
these off shell amplitudes are proportional to $p_i^2$ ($p_i$ is
the off-shell momentum) and they vanish when $p_i$ is also on
shell ($p_i^2=0$). We note that the vanishing for the $4$-particle
on-shell amplitudes were also obtained in \cite{GeorgiouKhoze} and
the googly amplitudes with 5 particles are also computed.

We relabel the off shell particle to be the first particle. The
helicity of this off-shell particle can be $1/2$, $-1/2$ (the
gluino) or $-1$ (the gluon). The corresponding off shell
amplitudes are:
\begin{eqnarray}
A_n(\Lambda_1^+,g_2^-,\cdots,\Lambda_s^-,\cdots,g_n^-) & = &
{\phi_s\over \phi_2\phi_n} {p_1^2 \over \prod_{i=2}^{n-1}[i,i+1]},
\\
 A_n(\Lambda_1^-,g_2^-,\cdots,\Lambda_r^+,\cdots,g_n^-) & = &
 -{\phi_r^3 \over \phi_2 \phi_n} {p_1^2 \over
\prod_{i=2}^{n-1}[i,i+1]},
\end{eqnarray}
and
\begin{equation}
A_n(g_1^-,\cdots,\Lambda_r^+,\cdots,\Lambda_s^-,\cdots,g_n^-)=
{\phi_r^3\phi_s\over \phi_2\phi_n} {p_1^2 \over
\prod_{j=2}^{n-1}[i,i+1]}.\label{eqfg}
\end{equation}
respectively.  The proof of the above  formulas is  quite similar
to the proof for eqs.~(\ref{eq34}) and  (\ref{eq22}) given in
\cite{Zhu}. We will not repeat the proof here.

 \begin{figure}[ht]
    \epsfxsize=80mm%
    \hfill\epsfbox{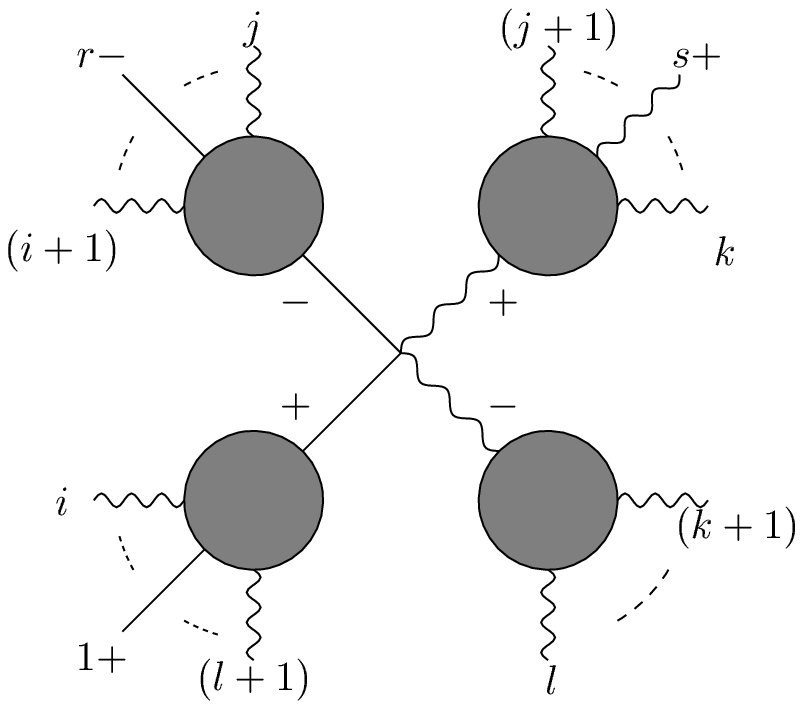}\hfill~\\
    \caption{The decomposition for the googly amplitude with
    fermions when $1\le i\le r-1, r\le j\le s-1,s\le k<l\le n$.
    }
    \label{figfg1}
   \end{figure}

 \begin{figure}[ht]
    \epsfxsize=80mm%
    \hfill\epsfbox{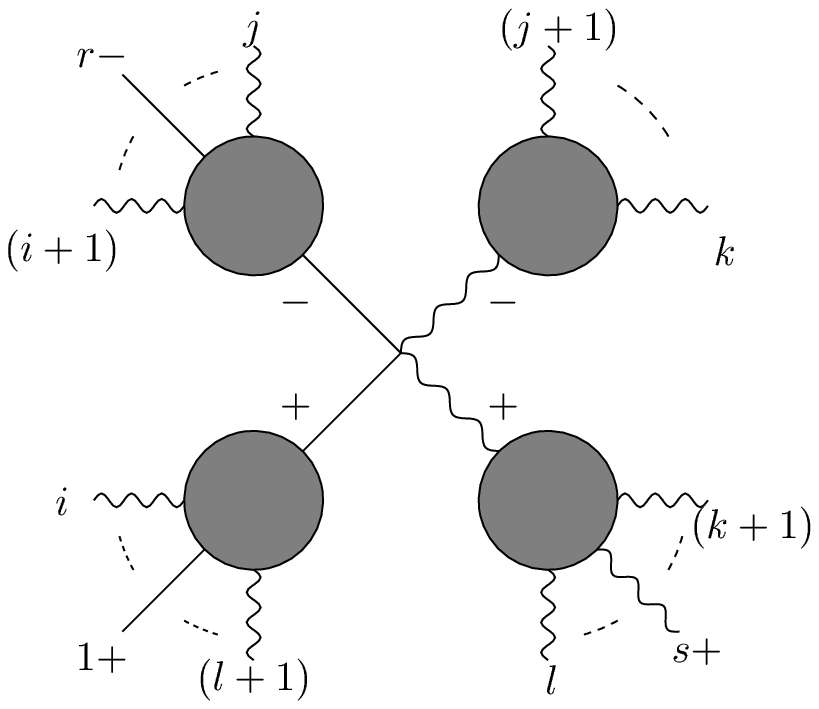}\hfill~\\
    \caption{The decomposition for the googly amplitude with
    fermions when $1\le i\le r-1,r \le j<k\le s-1, s\le l\le n$.
    }
    \label{figfg2}
   \end{figure}

 \begin{figure}[ht]
    \epsfxsize=100mm%
    \hfill\epsfbox{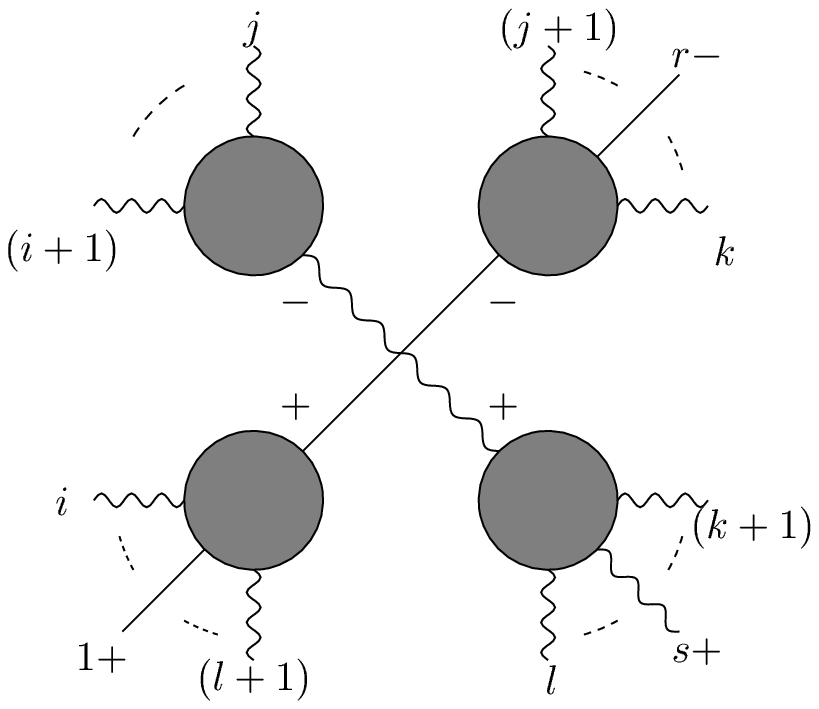}\hfill~\\
    \caption{The decomposition for the googly amplitude with
    fermions when $1\le i<j\le r-1,r \le k\le s-1, s\le l\le n$.
    }
    \label{figfg3}
   \end{figure}

 \begin{figure}[ht]
    \epsfxsize=80mm%
    \hfill\epsfbox{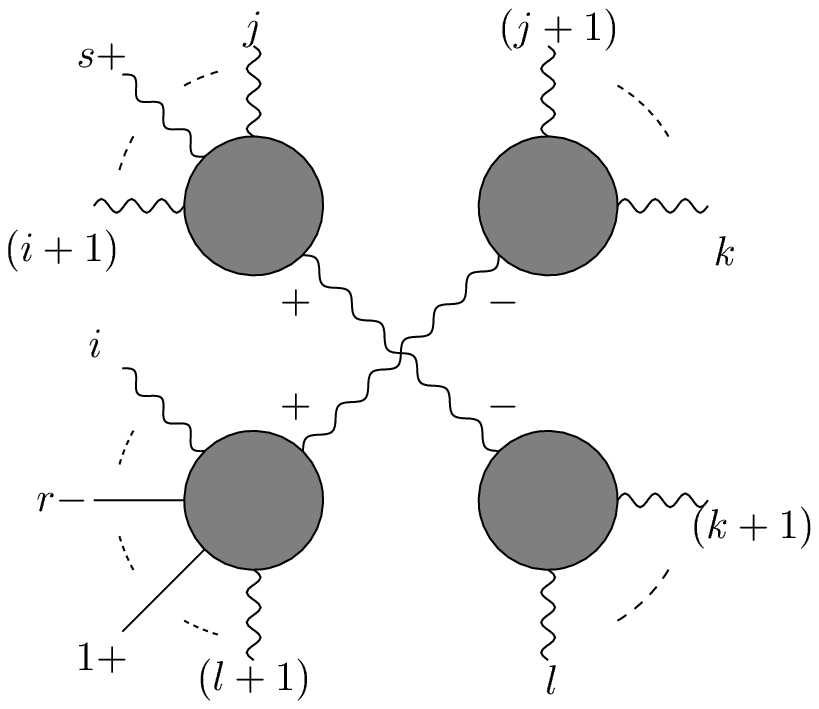}\hfill~\\
    \caption{The decomposition for the googly amplitude with
    fermions when $r\le i\le s-1,s \le j<k<l\le n$.
    }
    \label{figfg4}
   \end{figure}

 \begin{figure}[ht]
    \epsfxsize=80mm%
    \hfill\epsfbox{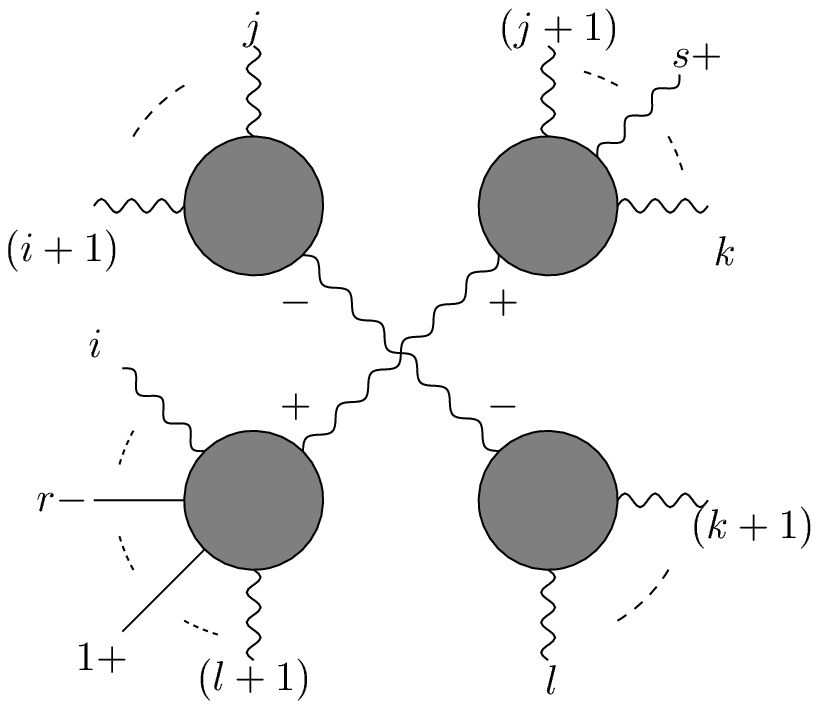}\hfill~\\
    \caption{The decomposition for the googly amplitude with
    fermions when $r\le i<j\le s-1,s \le k<l\le n$.
    }
    \label{figfg5}
   \end{figure}

 \begin{figure}[ht]
    \epsfxsize=80mm%
    \hfill\epsfbox{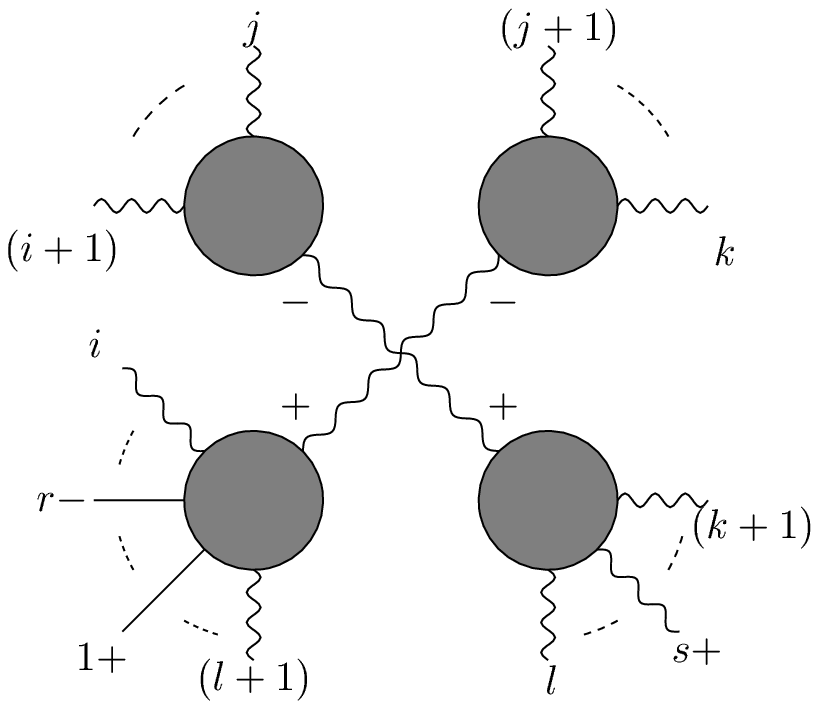}\hfill~\\
    \caption{The decomposition for the googly amplitude with
    fermions when $r\le i<j<k\le s-1,s<l\le n$.
    }
    \label{figfg6}
   \end{figure}

Now we  begin to calculate the googly amplitudes.  First we
consider the googly amplitudes with $2$ gluinos:
$A_n(\Lambda_1^+,g_2^-,\cdots,\Lambda_r^-, \cdots, g_s^+, \cdots,
g_n^- )$.  We will consider the case for $1<r<s \ne n$ only. The
rest cases are similar. As we proved in section 3,  there is only
one vertex with $4$ lines. So we can again use the same
organization of all the contributing diagrams as in the
computation of googly amplitude with one quark-antiquark pair in
section 4.

All contributing diagrams are shown in the figures from
Fig.~\ref{figfg1} to Fig.~\ref{figfg6}. The various contributions
corresponding to the Fig.~\ref{figfg1}, Fig.~\ref{figfg2},
Fig.~\ref{figfg3}, Fig.~\ref{figfg4}, Fig.~\ref{figfg5} and
Fig.~\ref{figfg6} are
\begin{eqnarray}
A_n^1&=&{\phi_1^3 \phi_r \phi_s^4 \over
\prod_{s=1}^n[s,s+1]}\sum_{i=1}^{r-1}\sum_{j=r}^{s-1}\sum_{k=s}^{n-1}\sum_{l=k+1}^n
{[i,i+1]\over \phi_i \phi_{i+1}} \,
{[k,k+1]\over \phi_k \phi_{k+1}} \,\nonumber \\
& & \times{[j,j+1]\over \phi_j \phi_{j+1}} \,{[l,l+1]\over \phi_l
\phi_{l+1}} \,  {-\lan V_4, V_2\ran^3 \over \lan V_1, V_2 \ran
\lan V_2, V_3\ran \lan V_3, V_4\ran},
\end{eqnarray}
\begin{eqnarray}
A_n^2&=&{\phi_1^3 \phi_r \phi_s^4 \over
\prod_{s=1}^n[s,s+1]}\sum_{i=1}^{r-1}\sum_{j=r}^{s-2}\sum_{k=j+1}^{s-1}\sum_{l=s}^n
{[i,i+1]\over \phi_i \phi_{i+1}} \,
{[k,k+1]\over \phi_k \phi_{k+1}} \,\nonumber \\
& & \times{[j,j+1]\over \phi_j \phi_{j+1}} \,{[l,l+1]\over \phi_l
\phi_{l+1}} \,  {\lan V_3, V_2\ran^2 \lan V_3, V_1 \ran\over \lan
V_1, V_2 \ran \lan V_3, V_4\ran \lan V_4, V_1\ran},
\end{eqnarray}
\begin{eqnarray}
A_n^3&=&{\phi_1^3 \phi_r \phi_s^4 \over
\prod_{s=1}^n[s,s+1]}\sum_{i=1}^{r-2} \sum_{j=i+1}^{r-1}
\sum_{k=r}^{s-1} \sum_{l=s}^n {[i,i+1]\over \phi_i \phi_{i+1}} \,
{[k,k+1]\over \phi_k \phi_{k+1}} \,\nonumber \\
& & \times{[j,j+1]\over \phi_j \phi_{j+1}} \,{[l,l+1]\over \phi_l
\phi_{l+1}} \,  {\lan V_2, V_3\ran^2 \over  \lan V_3, V_4\ran \lan
V_4, V_1\ran},
\end{eqnarray}
\begin{eqnarray}
A_n^4&=&{\phi_1^3 \phi_r \phi_s^4 \over
\prod_{s=1}^n[s,s+1]}\sum_{i=r}^{s-1} \sum_{j=s}^{n-2}
\sum_{k=j+1}^{n-1} \sum_{l=k+1}^n {[i,i+1]\over \phi_i \phi_{i+1}}
\,
{[k,k+1]\over \phi_k \phi_{k+1}} \,\nonumber \\
& & \times{[j,j+1]\over \phi_j \phi_{j+1}} \,{[l,l+1]\over \phi_l
\phi_{l+1}} \,  {\lan V_3, V_4\ran^3 \over \lan V_1, V_2 \ran \lan
V_2, V_3\ran \lan V_4, V_1\ran},
\end{eqnarray}
\begin{eqnarray}
A_n^5&=&{\phi_1^3 \phi_r \phi_s^4 \over
\prod_{s=1}^n[s,s+1]}\sum_{i=r}^{s-2} \sum_{j=i+1}^{s-1}
\sum_{k=s}^{n-1} \sum_{l=k+1}^n {[i,i+1]\over \phi_i \phi_{i+1}}
\,
{[k,k+1]\over \phi_k \phi_{k+1}} \,\nonumber \\
& & \times{[j,j+1]\over \phi_j \phi_{j+1}} \,{[l,l+1]\over \phi_l
\phi_{l+1}} \,  {\lan V_2, V_4\ran^4 \over \lan V_1, V_2 \ran \lan
V_2, V_3\ran \lan V_3, V_4 \ran \lan V_4, V_1\ran},
\end{eqnarray}
and
\begin{eqnarray}
A_n^6&=&{\phi_1^3 \phi_r \phi_s^4 \over
\prod_{s=1}^n[s,s+1]}\sum_{i=r}^{s-3} \sum_{j=i+1}^{s-2}
\sum_{k=j+1}^{s-1} \sum_{l=s}^n {[i,i+1]\over \phi_i \phi_{i+1}}
\,
{[k,k+1]\over \phi_k \phi_{k+1}} \,\nonumber \\
& & \times{[j,j+1]\over \phi_j \phi_{j+1}} \,{[l,l+1]\over \phi_l
\phi_{l+1}} \,  {\lan V_2, V_3\ran^3 \over \lan V_1, V_2 \ran \lan
V_3, V_4\ran \lan V_4, V_1\ran},
\end{eqnarray}
where\footnote{Here and in the following, the summation like
$\sum_{s=l+1}^{n+i}$ is understood as
$\sum_{s=l+1}^n+\sum_{s=1}^i$.}
\begin{eqnarray}
V_1 = \sum_{s=l+1}^{n+i} \lambda_s \phi_s , \qquad & &
V_2 = \sum_{s=i+1}^j \lambda_s \phi_s , \label{eqvii} \\
V_3 = \sum_{s=j+1}^k \lambda_s \phi_s , \qquad & &
 V_4 =\sum_{s=k+1}^l \lambda_s \phi_s ,\label{eqvi}
\end{eqnarray}

The sum of   $A_n^{1,2,3}$ can be written as
\begin{eqnarray}
A_n^1+A_n^2+A_n^3&=&{\phi_1^3 \phi_r \phi_s^4 \over
\prod_{s=1}^n[s,s+1]}\sum_{i=1}^{r-1} \, \sum_{l={\rm
 max}\{s,i+3\}}^n\, \sum_{j=i+1}^{l-2}
\sum_{k=j+1}^{l-1} \nonumber \\
& & \times {[i,i+1]\over \phi_i \phi_{i+1}} \, {[k,k+1]\over
\phi_k \phi_{k+1}} \, {[j,j+1]\over \phi_j \phi_{j+1}}
\,{[l,l+1]\over \phi_l
\phi_{l+1}} \,  \nonumber \\
& &\times{-\lan V_a, V_b\ran^3 \lan V_a,V_1\ran \over \lan V_1,
V_2 \ran \lan V_2, V_3\ran \lan V_3, V_4\ran \lan V_4,
V_1\ran},\label{eq123}
\end{eqnarray}
where $a$   is an index ($a=3,4$) which satisfies the condition
that $V_a$ includes neither $\lambda_r\phi_r$ nor
$\lambda_s\phi_s$ as defined in eqs.~(\ref{eqvii}) and
(\ref{eqvi}),  and $b$ is an index ($b=2,3,b<a$) which satisfies
the condition that $V_b$ includes $\Lambda_s\phi_s$ as defined in
eqs.~(\ref{eqvii}) and (\ref{eqvi}).

The sum of $A_n^{4,5,6}$ can be similarly organized and we have:
\begin{eqnarray}
A_n^4+A_n^5+A_n^6&=&{\phi_1^3 \phi_r \phi_s^4 \over
\prod_{s=1}^n[s,s+1]}\sum_{i=r}^{s-1}\, \sum_{l={\rm
 max} \{s,i+3\}}^n\, \sum_{j=i+1}^{l-2}
\sum_{k=j+1}^{l-1} \nonumber \\
& & \times {[i,i+1]\over \phi_i \phi_{i+1}} \, {[k,k+1]\over
\phi_k \phi_{k+1}} \,  {[j,j+1]\over \phi_j \phi_{j+1}}
\,{[l,l+1]\over \phi_l \phi_{l+1}} \,  \nonumber \\
&&\times{\lan V_p, V_q\ran^4 \over \lan V_1, V_2 \ran \lan V_2,
V_3\ran \lan V_3, V_4\ran \lan V_4, V_1\ran}, \label{eq456}
\end{eqnarray}
where $p$ and $q$    are the two indexes ($p=2,3$, $q=3,4$, $p\ne
q$) which satisfy that neither $V_p$ nor $V_q$ includes
$\lambda_r\phi_r$ as defined in eqs.~(\ref{eqvii}) and
(\ref{eqvi}).

In order to obtain the correct result for the googly amplitude,
all we need is to prove the following identity:
 \begin{eqnarray}
& &\hskip -1cm \sum_{i=1}^{r-1}\,\sum_{l={\rm max}\{s,i+3\}}^n\,
 \, \sum_{j=i+1}^{l-2} \sum_{k=j+1}^{l-1}\,
{[i, i+1] \over \phi_i\phi_{i+1}}{[j, j+1] \over \phi_j\phi_{j+1}}\nonumber\\
& & \hskip -.5cm \times{[k, k+1] \over \phi_k\phi_{k+1}} {[l, l+1]
\over \phi_l\phi_{l+1}} {-\lan V_a, V_b\ran^3 \lan V_a,V_1\ran
\over \lan V_1, V_2 \ran \lan V_2, V_3\ran \lan V_3, V_4\ran
\lan V_4, V_1\ran}\nonumber\\
& & \hskip -.5cm +\sum_{i=r}^{s-1}\,\sum_{l= {\rm
max}\{s,i+3\}}^n\, \sum_{j=i+1}^{l-2}
\sum_{k=j+1}^{l-1} {[i, i+1] \over \phi_i\phi_{i+1}}{[j, j+1] \over \phi_j\phi_{j+1}}
\nonumber\\
& & \hskip -.5cm  \times{[k, k+1] \over \phi_k\phi_{k+1}} {[l,
l+1] \over \phi_l\phi_{l+1}} {\lan V_p, V_q\ran^4 \over \lan V_1,
V_2 \ran \lan V_2, V_3\ran \lan V_3, V_4\ran \lan
V_4, V_1\ran} \nonumber\\
&=&{[s,1]^3[s,r]\over \phi_s^4 \phi_1^3 \phi_r}.
\end{eqnarray}
This is indeed the case and one can prove it by using the same
method used in \cite{Zhu,Wu} and appendix A. We will not give the
proof here.

By using this identity , we have:
 \begin{equation}
A_n=\sum_{i=1}^6 A_n^i=[s,1]^3 [s,r]/(\prod_{i=1}^n[i,i+1]),
\end{equation}
which is the right result for the googly amplitude with 2 gluinos.

As we did in section, we can similarly calculate the googly
amplitudes with $4$-gluinos (all gluons then have negative
helicity) and the result is:
\begin{equation}
A  (g_1^-, \cdots,\Lambda_r^+,\cdots,\Lambda_s^+,
\cdots,\Lambda_p^-, \cdots,\Lambda_q^-,\cdots, g_n^-)=
-{[p,q]^3[r,s]\over \prod_{i=1}^n[ i,i+1]} ,
\end{equation}
as expected. We will not present any details here.

\section*{Acknowledgments}

We would like to thank Zhe Chang, Bin Chen, Han-Ying Guo, Miao Li,
Jian-Xin Lu, Jian-Ping Ma, Ke Wu and Yong-Shi Wu  for discussions.
Jun-Bao Wu would like to thank Qiang Li for help on drawing the
figures. Chuan-Jie Zhu would like to thank Jian-Xin Lu  and the
hospitality at the Interdisciplinary Center for Theoretical Study,
University of Science and Technology of China where part of this
work was done.

\section*{Appendix A: The proof of eq.~(\ref{eqidf}) }

In    this appendix,    we will prove   eq.~(\ref{eqidf}). As in
\cite{Zhu, Wu}, we can use an $SL(2,{\bf C})$ transformation and a
rescaling of $\tilde\eta$ to choose $\tilde\eta^1=0$ and
$\tilde\eta^2=1$. We then do a rescaling of $\tilde\lambda_{i1}$
by $\tilde\lambda_{i2}$, i.e. by defining $\vp_i =
{\tilde\lambda_{i1} \over \tilde\lambda_{i2}} $, and also do a
rescaling of $\lambda_{ia}$ by $1/\tilde\lambda_{i2}$. After using
these redefinitions,  Eq.~(\ref{eqidf}) becomes:
\begin{eqnarray}
& &\hskip -.5cm F_f(\vp_i) \equiv (\vp_{\bar
q}-\vp_q)\sum_{i=0}^{I-1} \sum_{k={\rm max}\{I,
i+2\}}^n\,\sum_{j=i+1}^{k-1}\,
(\vp_i-\vp_{i+1}) (\vp_j-\vp_{j+1}) \nonumber \\
&& \times(\vp_k-\vp_{k+1}) {-\lan V_1, V_p\ran \lan V_4, V_p
\ran^3\over \lan V_1, V_2 \ran \lan V_2, V_3\ran \lan V_3, V_4
\ran
\lan V_4, V_1\ran}\nonumber \\
&& +\sum_{i=0}^{I-1}\,\sum_{l={\rm max}\{I,
i+3\}}^n\,\sum_{j=i+1}^{l-2}\sum_{k=j+1}^{l-1} (\vp_i-\vp_{i+1})
(\vp_j-\vp_{j+1}) (\vp_k-\vp_{k+1}) (\vp_l-\vp_{l+1}) \nonumber \\
& & \times{\lan \tilde V_r,\tilde V_s \ran^4\over \lan \tilde V_1,
\tilde V_2 \ran \lan \tilde V_2, \tilde V_3\ran \lan \tilde
V_3,\tilde V_4 \ran\lan \tilde V_4,\tilde V_1\ran}=(\vp_q-\vp_I)^3
(\vp_{\bar q}-\vp_I),\label{eqf27}
\end{eqnarray}
where
\begin{eqnarray}
&& V_1=\sum_{s=0}^i \lambda_s,  \quad V_2=\sum_{s=i+1}^j
\lambda_s,\\
&& V_3=\sum_{s=j+1}^k \lambda_s,  \quad V_4=\sum_{s=k+1}^{n+1}
\lambda_s. \\
&& \tilde V_1=\sum_{s=0}^i \lambda_s +\sum_{s=l+1}^{n+1}
\lambda_s,
  \quad \tilde V_2=\sum_{s=i+1}^j \lambda_s, \\
&& \tilde V_3=\sum_{s=j+1}^k \lambda_s,  \quad \tilde
V_4=\sum_{s=k+1}^l \lambda_s,
\end{eqnarray}
There are also two constraints:
\begin{eqnarray}
 & & \sum_{i=1}^4 V_i=\sum_{i=1}^4
 \tilde V_i= \sum_{l=0}^{n+1} \lambda_l = 0 ,
 \label{eqf29} \\
 & & \sum_{l= 0}^{n+1} \lambda_i\,  \varphi_i = 0 , \label{eqf30}
 \end{eqnarray}
from momentum conservation.

From eq.~(\ref{eqf29}) and eq.~(\ref{eqf30}) we can solve
$\lambda_0$ and $\lambda_I$ in terms of the rest $\lambda_{i}$ and
all $\varphi_j$'s  as:
\begin{eqnarray}
&& \lambda_0=-\sum_{1\le j\le n+1, j \ne
I}{\vp_j-\vp_I\over\vp_0-\vp_I}\lambda_j,\label{eqla1}\\
&&\lambda_r=\sum_{1\le j\le n+1, j \ne
I}{\vp_j-\vp_0\over\vp_0-\vp_I}\lambda_j.\label{eqla2}
\end{eqnarray}
By using this solution, we can consider  $F_f(\vp_i)$ as a
function of $\lambda_{j}$ ($1\le j\le n+1$, $j\ne I$) and all
$\varphi_j$'s ($0\le j\le n+1$).

To prove eq.~(\ref{eqf27}), we first prove that  $F_f(\vp_i)$ is
independent of $\varphi_j$ for $1\le j\le n, j\ne I$ and
$F_f(\vp_i)$ depends on $\vp_{n+1}$ linearly by analyzing the pole
terms.

First we will show that when $\vp_s \to \infty (1\le s \le n, s
\ne I)$ the pole terms are vanishing in $F_f(\vp)$ and when
$\vp_{n+1}\to\infty$, $F_f(\vp)$ will grow as $\vp_{n+1}$ at most.

When $\vp_s \to \infty (1\le s \le n, s \ne I)$,   $\lambda_1$ and
$\lambda_I$ will grow as $\vp_s$, and other $\lambda_i$'s don't
depend on $\vp_s$. One can easily find that the pole terms in the
second sum in eq.~(\ref{eqf27}),
\begin{eqnarray}
&&\sum_{i=0}^{I-1}\,\sum_{l={\rm max}\{I,
i+3\}}^n\,\sum_{j=i+1}^{l-2}\sum_{k=j+1}^{l-1} (\vp_i-\vp_{i+1})
(\vp_j-\vp_{j+1}) (\vp_k-\vp_{k+1}) (\vp_l-\vp_{l+1}) \nonumber \\
& & \hskip 2cm \times{\lan \tilde V_r,\tilde V_s \ran^4\over \lan
\tilde V_1, \tilde V_2 \ran \lan \tilde V_2, \tilde V_3\ran \lan
\tilde V_3,\tilde V_4 \ran\lan \tilde V_4,\tilde
V_1\ran},\label{eqf28}
\end{eqnarray}
are vanishing as in \cite{Zhu, Wu}.

As to  the terms  in the first sum in eq.~(\ref{eqf27}), it is
easy to see that $V_1$ and $V_q$ ($q=2,3, q\ne p$) will grow as
$\vp_s$.

If none of the $V_i$'s ($i=1,\cdots,4$) includes only the term
$\lambda_s$, the factor $F(i,j,k)\equiv(\vp_{\bar
q}-\vp_q)(\vp_i-\vp_{i+1}) (\vp_j-\vp_{j+1}) (\vp_k-\vp_{k+1})$
grows as $\vp_s$ at most, $-\lan V_1, V_p\ran \lan V_4, V_p
\ran^3$ grow as $\vp_s$ at most, and $\lan V_1, V_2\ran \lan V_2,
V_3\ran \lan V_3, V_4\ran \lan V_4, V_1\ran$ grows as $\vp_s^3$ at
least. So the pole terms are vanishing.

The other case is when one of the $V_i$'s includes only the term
$\lambda_s$ (this $i$ must be $p$), then $F(i,j,k)$ grows as
$\vp_s^2$ at most, $-\lan V_1, V_p\ran \lan V_4, V_p \ran^3$ is
independent of $\vp_s$, and $\lan V_1, V_2\ran \lan V_2, V_3\ran
\lan V_3, V_4\ran \lan V_4, V_1\ran$ grows as $\vp_s^2$ at least.
So no terms will tend to infinity.

So when $\vp_s \to \infty$ $(1\le s \le n, s \ne I)$ the pole
terms are vanishing.

When   $\vp_{n+1}\to\infty$, $\lambda_1$ and $\lambda_I$ will grow
as $\vp_{n+1}$, and other $\lambda_i$'s don't depend on
$\vp_{n+1}$. One can find that the pole terms in
eq.~(\ref{eqf28}), are still vanishing.

As to the terms in the first sum in eq.~(\ref{eqf27}), it is easy
to see that $V_1$ and $V_q$ ($q=2,3, q\ne p$) will grow as
$\vp_{n+1}$.

If none of the $V_i$'s ($i=1,\cdots,4$) includes only the term
$\vp_{n+1}$, the factor $F(i,j,k)$ grows as $\vp_{n+1}$ at most,
$-\lan V_1, V_p\ran \lan V_4, V_p \ran^3$ grow as $\vp_{n+1}$ at
most, and $\lan V_1, V_2\ran \lan V_2, V_3\ran \lan V_3, V_4\ran
\lan V_4, V_1\ran$ grows as $\vp_{n+1}^3$ at least. So the pole
terms are vanishing.

The other case is when one of the $V_i$'s includes only the term
$\vp_{n+1}$ (this $i$ must be $4$), then  $F(i,j,k)$ grows as
$\vp_{n+1}^2$ at most, $-\lan V_1, V_p\ran \lan V_4, V_p \ran^3$
grows as $\vp_{n+1}$ as most, and $\lan V_1, V_2\ran \lan V_2,
V_3\ran \lan V_3, V_4\ran \lan V_4, V_1\ran$ grows as
$\vp_{n+1}^2$ at least. So every term grows as $\vp_{n+1}$ at
most.

So when $\vp_{n+1}\to\infty$, $F_f(\vp)$ will grow as $\vp_{n+1}$
at most.

Now we will show that the finite poles terms will vanishing.

There are pole terms   if any factor of $\langle V_1, V_2 \rangle
\lan V_2, V_3 \ran\langle V_3, V_4 \rangle \langle V_4, V_1
\rangle $ is vanishing or any factor of $\langle \tilde V_1,
\tilde V_2 \rangle \lan \tilde V_2, \tilde V_3 \ran \langle \tilde
V_3, \tilde V_4 \rangle \langle \tilde V_4, \tilde V_1 \rangle $
is vanishing. Let us consider first the vanishing of $\langle V_1,
V_2 \rangle$. We denote this set of $V_1$ and $V_2$ as $v_1$ and
$v_2$:
\begin{equation}
v_1 = \lambda_0 + \cdots + \lambda_{n_1}, \qquad v_2 =
\lambda_{n_1+1} + \cdots + \lambda_{n_2}.
\end{equation}

 \begin{figure}[ht]
    \epsfxsize=80mm%
    \hfill\epsfbox{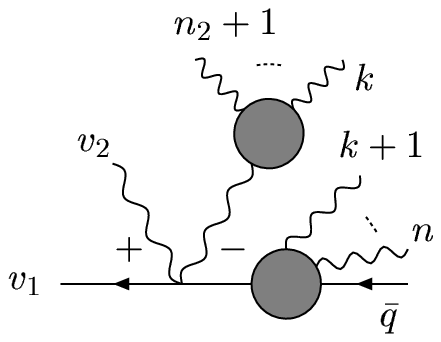}\hfill~\\
    \caption{Contributing diagrams from the vanishing of
    $\langle  V_1,  V_2 \rangle$. The other 2 contributing diagrams
    from the vanishing of  $\langle  V_3,  V_4 \rangle$ or
    $\langle  V_4,  V_1 \rangle$ are similar and are not shown.}
    \label{quarkpole1}
   \end{figure}

If $n_2<I$, then every  individual term will not tend to infinity.
So we turn to consider the case when $n_1<I\le n_2$. As one can
see from Fig.~\ref{quarkpole1} that there are contributions from
summing over $k$ and fixing $i=n_1$ and $j=n_2$. The residues
(ignoring an overall factor $(\vp_{\bar q}-\vp_q)(\varphi_{n_1}-
\varphi_{n_1+1}) (\varphi_{n_2}- \varphi_{n_2+1})$)  for these
pole terms are:
\begin{equation}
C_1 =  \sum_{k=n_2 + 1}^n \,(\varphi_k - \varphi_{k+1})\,  {-
\langle  V_3,V_4\rangle^2 \lan V_3, v_1\rangle \over \langle
v_2,V_3 \rangle \langle V_4,v_1 \rangle } \, .
\end{equation}
Because $\langle v_1, v_2 \rangle=0$, $v_1$ and $v_2$ are linearly
dependent, we can assume that $v_i=\a_i v_0, i=1,2$, for some
$\a_i$ and $v_0$.

Using this result and $v_1+v_2+V_3+V_4=0$, we have
\begin{eqnarray}
C_1={(\a_1+\a_2)^2 \over
\a_2}\sum_{k=n_2+1}^n(\vp_k-\vp_{k+1})\langle v_0, V_3 \rangle.
\label{eqc1}
\end{eqnarray}

Similar pole terms can also be obtained from the vanishing of the
factor $\langle V_3,V_4\rangle$  by setting $V_3 = v_2$ and $V_4 =
-(v_1 + v_2)$.  This gives the following contribution:
\begin{eqnarray}
C_2 & = &  \sum_{i= 0}^{n_1-1} \,(\varphi_i - \varphi_{i+1})\,  {
\langle  V_2,-v_1-v_2\rangle^3 \over \langle V_2, v_2 \rangle
\langle -(v_1+v_2),V_1 \rangle  } \,
\nonumber \\
& = & {(\a_1+\a_2)^2 \over \a_2}\sum_{i= 0}^{n_1-1} \,(\varphi_i-
\varphi_{i+1})\,
   \langle  v_0,V_1\rangle. \end{eqnarray}
The contribution obtained from    the vanishing of the factor
$\langle V_4,V_1\rangle$  by setting $V_1 = v_1$ and $V_4 = -(v_1
+ v_2)$   is:
\begin{equation}
C_3    =  {(\a_1+\a_2)^2 \over \a_2}\sum_{j= n_1+1}^{n_2-1}
\,(\varphi_j- \varphi_{j+1})\,
   \langle  v_0,V_1\rangle .
 \end{equation}
Using the same algebraic manipulation  as given  in \cite{Zhu,
Wu}, one   finds that $C_1+C_2+C_3=0$.

Now we consider the case when $\lan \tilde V_1, \tilde V_2\ran$
vanishing. We denote this set of $\tilde V_1$ and $\tilde V_2$ as
$\tilde v_1$ and $\tilde v_2$:
\begin{eqnarray}
&&\tilde v_1 =\lambda_{n_3+1} +\cdots +\lambda_n+\lambda_{\bar
q}+\lambda_q+\lambda_1 + \cdots + \lambda_{n_1}, \\
&& \tilde v_2 = \lambda_{n_1+1} + \cdots + \lambda_{n_2}.
\end{eqnarray}
We consider the case for $n_1+1\le I\le n_2$, the case for
$n_2+1\le I\le n_3$ can be treated similarly. The rest cases are
more easier as in \cite{Wu}.

 \begin{figure}[ht]
    \epsfxsize=70mm%
    \hfill\epsfbox{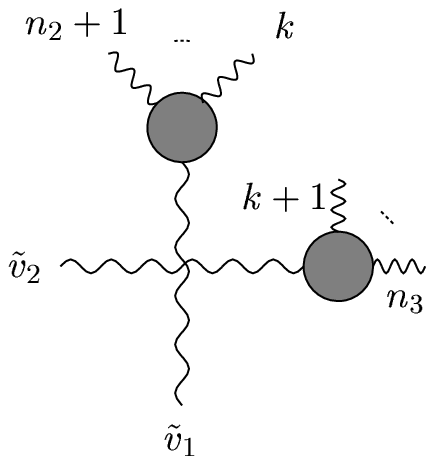}\hfill~\\
    \caption{Contributing diagrams from the vanishing of
    $\langle  \tilde V_1,  \tilde V_2 \rangle$. }
        \label{quarkpole4}
   \end{figure}

 \begin{figure}[ht]
    \epsfxsize=80mm%
    \hfill\epsfbox{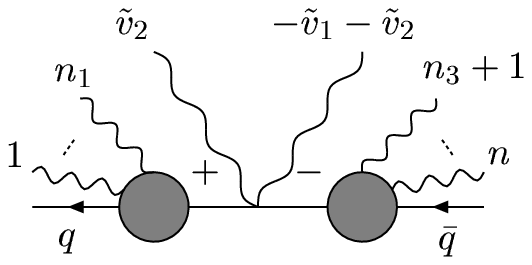}\hfill~\\
    \caption{Contributing diagrams from the vanishing of
    $\langle  V_2,  V_3 \rangle$. }
    \label{quarkpole9}
   \end{figure}

The residues corresponding to   Fig.~\ref{quarkpole4} from the
vanishing of $\langle \tilde V_1, \tilde V_2\rangle$ is:
\begin{eqnarray}\tilde C_1&=&\sum_{k=n_2 + 1}^{n_3-1}
\,(\varphi_k - \varphi_{k+1})\,  { \langle \tilde V_3,-\tilde
v_1-\tilde v_2-\tilde V_3 \rangle^3 \over \langle \tilde
v_2,\tilde V_3 \rangle \langle -\tilde v_1-\tilde v_2-\tilde
V_3,v_1 \rangle }\,
\nonumber \\
& = & {(\a_1+\a_2)^3 \over \a_1 \a_2}\sum_{k= n_2+1}^{n_3-1}
\,(\varphi_k- \varphi_{k+1})\,
   \langle \tilde v_0, \tilde V_3\rangle, \label{eqfactor}
\end{eqnarray}
by omitting an overall factor $(\varphi_{n_3}-
\varphi_{n_3+1})(\varphi_{n_1}- \varphi_{n_1+1}) (\varphi_{n_2}-
\varphi_{n_2+1})$. The other contributions from the vanishing of
the various factors $\langle \tilde V_4, \tilde V_1\rangle$ (2
contributions depending on the position of the positive helicity
gluon $g_I$), $\langle \tilde V_3, \tilde V_4\rangle$, $\langle
\tilde V_2, \tilde V_3\rangle$. We will not shown the diagrams
here.  Their respective contributions are:
\begin{eqnarray}
\tilde C_2 & = & {(\a_1+\a_2)^3 \over \a_1 \a_2}\sum_{j=
I}^{n_2-1} \,(\varphi_j- \varphi_{j+1})\,
   \langle \tilde v_0, \tilde V_2\rangle,
\\
\tilde C_3 & = & {(\a_1+\a_2)^3 \over \a_1 \a_2}\sum_{j=
n_1+1}^{I-1} \,(\varphi_j- \varphi_{j+1})\,
   \langle \tilde v_0, \tilde V_2\rangle ,
\\
\tilde C_4 & = & {(\a_1+\a_2)^3 \over \a_1 \a_2}\sum_{i=
1}^{n_1-1} \,(\varphi_i- \varphi_{i+1})\,
   \langle \tilde v_0, \tilde V_1\rangle , \\
\tilde C_5 & = & {(\a_1+\a_2)^3 \over \a_1 \a_2}\sum_{l= n_3+1}^n
\,(\varphi_l- \varphi_{l+1})\,
   \langle \tilde v_0, \tilde V_1\rangle,
\end{eqnarray}
by omitting the same overall factor as in eq.~(\ref{eqfactor}).
  The last contribution is from
the vanishing of the factor  $\langle  V_2, V_3\rangle$ as shown
in Fig.~\ref{quarkpole9}. It contribution is:
\begin{eqnarray}
\tilde C_6 & = & (\varphi_{\bar q} - \varphi_q)\,  { -\langle
-\tilde v_1- \tilde v_2, \tilde V_4 \rangle^2 \lan
 -\tilde v_1-\tilde v_2,\tilde v_1-\tilde V_4\ran \over \langle \tilde v_1-\tilde V_4,\tilde
v_2 \rangle \langle \tilde V_4,\tilde v_1-\tilde V_4 \rangle }\,
\nonumber \\
& = & {(\a_1+\a_2)^3 \over \a_1 \a_2} \,(\varphi_{\bar q}-
\varphi_q)\,
   \langle \tilde v_0, \tilde V_4\rangle .
   \end{eqnarray}

By using all the above results,  we have:
\begin{equation}
\sum_{i=1}^6 \tilde C_i=0.
\end{equation}
This shows that there is no finite pole terms.

Summarizing the above results from analyzing all the pole terms,
we conclude that
\begin{equation}
F_f(\vp)=a_1 \vp_{n+1}+a_2,
\end{equation}
where $a_1$ and $a_2$ are functions of $\vp_0,\vp_I$ and
$\lambda_s, 1 \le s\le n+1, s\le I$. Then we can fixed $a_1$ and
$a_2$ by a special set of $\varphi_s,1 \le s\le n+1, s\le I$.

We can make a convenient choice such as:
$\vp_2=\cdots=\vp_{s-1}=x, \vp_{s+1}=\cdots=\vp_{n+1}=y$. After
doing some algebras as  in \cite{Zhu, Wu}, we have
\begin{equation}
F_f(\vp)=(\vp_q-\vp_I)^3 (y-\vp_I).
\end{equation}
Since $y$ can   take any value, we must have:
\begin{equation}
a_1=(\vp_q-\vp_I)^3, \qquad a_2=-\vp_I (\vp_q-\vp_I)^3,
\end{equation}
and so $F_f(\vp)=(\vp_q-\vp_I)^3 (\vp_{\bar q}-\vp_I)$. This ends
our proof of eq.~(\ref{eqidf}).

\end{document}